\documentclass[apj,numberedappendix]{emulateapj}
\usepackage[]{natbib}
\usepackage{psfig,graphicx,ulem,epsfig}
\usepackage{txfonts}
\usepackage{soul,color}
\usepackage{longtable}

\usepackage{color}

\newcommand{\simgt}{\lower.5ex\hbox{$\; \buildrel > \over \sim \;$}}
\newcommand{\simlt}{\lower.5ex\hbox{$\; \buildrel < \over \sim \;$}}

\shorttitle{Star formation activity in a young galaxy cluster at $z=0.866$} \shortauthors{Lagan\'{a} et al.}

\begin{document}

\title{Star formation activity in a young galaxy cluster at $z=0.866$}

\author{T. F. Lagan\'{a}\altaffilmark{1}} 
\author{M.~P.~Ulmer\altaffilmark{2}} 
\author{L.~P. Martins\altaffilmark{1}} 
\author{E. da Cunha\altaffilmark{3}} 

\altaffiltext{1}{NAT, Universidade Cruzeiro do Sul, Rua Galv\~{a}o Bueno, 868, CEP: 01506-000,S\~ao Paulo/SP, Brazil}
\altaffiltext{2}{Department of Physics and Astronomy and CIERA, Northwestern University, 2145 Sheridan Road, Evanston IL 60208-3112, USA}
\altaffiltext{3}{Center for Astrophysics and Supercomputing, Swinburne University of Technology, Hawthorn VIC 3122, Australia}


\begin{abstract}

The galaxy cluster RXJ1257$+$4738 at $z=0.866$ is one of the highest redshift clusters with a richness of 
multi-wavelength data, and thus a good target to study the star formation-density relation at early epochs.
Using a sample of spectroscopically-confirmed cluster members,
we derive the star formation rates of our galaxies using two methods, 
(I) the relation between SFR and total infrared luminosity 
extrapolated from the observed \textit{Spitzer} MIPS 24$\mu$m imaging data, and (II) spectral 
energy distribution (SED) fitting
using the MAGPHYS code, including eight different bands.

We show that, for this cluster, the SFR-density relation is very weak and seems to be dominated by the 
two central galaxies and the SFR presents a mild dependence on stellar mass, with more massive galaxies
having higher SFR. However, the specific SFR (SSFR) decreases with stellar mass, meaning that more massive 
galaxies are forming less stars per unit of mass, and thus suggesting that the increase in star-forming 
members is driven by cluster assembly and infall.

 If the environment is somehow driving the SF, one would expect a relation between the SSFR and the
cluster centric distance, but that is not the case. A possible scenario to explain this lack of 
correlation is the contamination by infalling  galaxies in the 18
inner part of the cluster which may be on their initial pass through the cluster center.
As these galaxies have higher SFRs for their stellar mass, they enhance the mean SSFR in the center of the cluster.

\end{abstract}


\maketitle

\section{Introduction}
\label{intro}
It has been known since the piooneering work of \citet{Dressler80}
that higher density environments in the local Universe host more elliptical and S0 galaxies, meaning that they are
dominated by evolved galaxy 
populations. This fact illustrates the
impact of the large-scale environment on galaxy properties. Since there is a strong correlation between 
the Hubble type and the star formation rate \citep[SFR;][]{kennicutt98}
a correlation between the SFR and local galaxy density is expected and  has been observed in  recent studies \citep[e.g.][]{Gomez03,Balogh04,Blanton09}.
That is, the relative number of star forming galaxies and quiescent galaxies varies strongly with local density, showing that red, quiescent,
early-type galaxies represent a large proportion of the population in higher density regions at low redshift, such as the core of galaxy groups and clusters,
while blue, star-forming, late-type galaxies are predominant in lower density regions.

Extensive observational effort at low redshift have yielded a paradigm for galaxy  environmental dependencies, such that the densest regions at $z\simeq0$ harbour 
massive \citep{Kauffmann04}, red \citep{Balogh04}, early-type passive galaxies \citep{Dressler80}. At $z\simeq0$, several studies 
have found the mean SFRs of galaxies to be lower at higher densities such as in groups and clusters \citep[e.g.,][]{Gomez03,Kauffmann04}.
However, while it is well established that higher redshift 
clusters contain increased star-formation
activity compared to their local counterparts  \citep[e.g.,][]{BO78,Bai07,Saintonge08,Patel11,Webb13}, which is in line with the rapid decline of cosmic SFR since $z\simeq1$ \citep{Lilly96,Madau96,LeFloch05,Hopkins06},
there has not yet been a consensus on the SFR-density relation at this epoch.
 For example, using luminosity-limited samples, \citet{Elbaz07} and \citet{Cooper08} find that, at $z\simeq1$, 
the SFR-density relation reverses, such that galaxies in higher density environments have enhanced SFRs. 
However, for a stellar mass-limited sample at $z\simeq0.8$, \citet{Patel09} find that the SFR declines at higher densities, 
much like at $z\simeq0$. The difference between these studies could be the result
 of different sample selections probing very different environment densities \citep{Patel09,Patel11}, nevertheless this shows that more work is needed to
 understand how the effect of environment on galaxy evolution evolves with redshift.

The cosmic time around redshift  $z\simeq1$ 
is crucial to understand cluster and galaxy evolution because it is most likely the epoch of the first mature galaxy clusters \citep[e.g.,][]{Kravtsov12}.
This makes the redshift range between 0.8 and 1.0 
particularly interesting for comparing star formation histories of galaxies in clusters
and in the field, as well as studying the relationship between infalling galaxies and gas heating. 
The direction of the SFR-density relation
at $z\simeq1$ has implications for the role of the various physical processes that operate in different environments
regulating star formation (SF) and the evolution of galaxies. 
It appears that the physical mechanisms that relate galaxies with their environment 
are more complex than initially thought, and cannot
be reduced solely to ram-pressure stripping from the intra-cluster medium (ICM) of massive galaxy clusters \citep[see the discussion in][]{Balogh04}.
If, for example, a direct correlation between the SFR and local density 
is observed at $z\simeq1$, that would imply that the physical processes leading to the quenching of star formation
in high density environments produce fewer red,
non-star forming galaxies.

A widely used method to follow the build-up of galaxies through time is 
 to look into the evolution of their optical colors. Observed galaxy populations show a bimodal distribution
 in color-magnitude diagrams, which separate red, passively evolving galaxies from the blue, actively star-forming galaxies.
Recent studies have demonstrated that this bimodality was already
in place at $z\simeq1$, with red galaxies preferentially located  in the densest regions \citep[e.g.,][]{Cooper06,Cucciati06}.
However, the optical color of a galaxy is not straightforwardly linked
with its instantaneous SFR, since it is also affected by dust content, and the past star formation history of the galaxy.

A recent strategy to overcome these difficulties and understand the star formation rate-density relation is to directly measure the instantaneous
SFR as a function of redshift and environment using multi-wavelength diagnostics beyond solely optical colours.  
In a previous  work, \citet{Ulmer09} presented a multi-wavelength analysis of the cluster RX J1257$+$4738 using optical, 
near- and far-infrared data to determine its dynamical state, and to gain insights into its galaxy population properties.
However, this study is incomplete, because the {\it Spitzer} data were not sufficiently well calibrated to be analyzed
at the time of that study, which could
potentially affect the measurements of the SFR and stellar mass of the cluster galaxies.

The {\it Spitzer} archive now contains reliably calibrated photometric data for this cluster in all observed bands, which allows us to carry out a detailed analysis 
of the stellar mass content and star formation activity of this distant X-ray emitting young cluster of galaxies. 
In this paper, we use the deepest existing mid-infrared survey at  24$\mu$m,
with some of the largest spectroscopic completeness to analyse the relationship between the SFR and 
the specific SFR (sSFR) as a function of projected cluster centric
distance (as an indication of the density) in a galaxy cluster at  $z=0.866$: RX J1257$+$4738.

In principle, this approach might seem similar to that of \citet{PintosCastro13} (hereafter PC13), who also
performed a multi-wavelength analysis of the galaxies in this cluster. Our current study differs from PC13 in two crucial aspects.
First, our spectroscopic observations allow us to establish cluster membership of the galaxies more robustly than PC13, who used
photometric redshifts in their analysis. Second, we find that in PC13 there was a misconception
that there were no MIPS sources associated with the western cluster members.  
We discuss the properties of these ``missing sources'', and explain why PC13 were not able to associate these MIPS sources
with cluster members. Since we successfully identified  MIPS sources near the cluster center that are clusters members, 
we  provide a different interpretation from PC13 of the observations of RX J1257 by 
updating the work of \citep{Ulmer09} with robust constraints of the SFR and specific SFR of the galaxies enabled by
the {\it Spitzer} data.

The paper is organized as follows: in \S~\ref{data} we present the spectroscopic and imaging data used in this paper, 
in \S~\ref{sfr} we  describe the methods to determine the physical properties of the galaxies (SFRs and stellar masses).
We present our results in  \S~\ref{res}, the discussion in \S~\ref{disc}, and present our conclusions
in  \S~\ref{conc}. Throughout this paper, we assume a concordant $\Lambda$CDM cosmology with 
$\Omega_{\rm m} = 0.27$, $\Omega_{\rm \Lambda} = 0.7$, $H_{0}=71 \, \rm km \,s^{-1} Mpc^{-1}$, 
from which we compute the angular scale as 7.72 kpc/arcsec, a luminosity distance ($\rm D_{L}) = 5544.7$~Mpc and an angular distance $(D_{A})= 1592.4$~Mpc.  
This cluster is centered at  R.A.$=$12h 57m 12.2s, Dec=+47$^{o}$ 38$^{\prime}$ 06.5$^{\prime \prime}$ (J2000)
and the virial radius, based on the temperature of the ICM and using the formula from \citet{Hilton07} is $R_{vir} =1.05$~Mpc 
\citep{Ulmer09,PintosCastro13}. \citet{Ulmer09} inferred an X-ray mass of (1-5) $\times 10^{14} M_{\odot}$ and based on 
a velocity dispersion of 600 km/s, these authors obtained an independent mass estimate of 6.1 $\times 10^{14} M_{\odot}$.
All magnitudes are quoted in the AB system and confidence intervals correspond to the 68\% confidence level unless otherwise stated.

\section{Data analysis}
\label{data}

 \subsection{\textit{Spitzer} imaging}

We use IRAC \citep[Infrared Array Camera,][]{Fazio04}
 and MIPS \citep[Multiband Imaging Photometer for {\it Spitzer},][]{Rieke04}
data downloaded from the {\it Spitzer} Heritage Archive\footnote{http://irsa.ipac.caltech.edu/data/SPITZER/docs/spitzerdataarchives/sha/} (SHA). 
There were 20 IRAC frames with 100s integration time per frame, centered 
at each of the channels 3.6, 4.5, 5.8 and 8 $\mu$m, and 30s integration time MIPS frames, centered at 24 $\mu$m, 
resulting in a total of 9050s exposure time.
The calibrated mosaic images from the standard {\it Spitzer} pipeline showed satisfactory removal of instrumental artifacts
and were sufficiently clean for the IRAC data analysis. However, the MIPS 24$\mu$m pipeline mosaic exhibited artifacts and gradients, 
which were significantly reduced through a zodiacal light subtraction, a self-calibrating flat, and an improved overlap correction algorithm. 
We converted the original flux values  from MJy/sr to $\mu$Jy,
taking into account the pixel size to which these images were rebinned and using the conversion factor 
provided in the IRAC Handbook.

One of the major differences between the data used in this work and in PC13 is regarding the MIPS 24$\mu$m fluxes.
It appears that the data used by PC13 were not the latest dataset processed through the {\it Spitzer} pipeline, 
which can be found in the {\it Spitzer} Heritage Archive (SHA). 
We note that the faintest source in our catalogue has a MIPS 24$\mu$m flux of 20~$\mu$Jy (i.e. $m=20.6$), which
is fainter than the completeness limit of $m=19.2$ quoted by PC13.
The MIPS flux density is measured using SHA with a default aperture of diameter 14.7 arcsec and checked using flux integration package 
(Aper referenced by the SHA) by hand to confirm if SHA  returns the correct fluxes of our targeted galaxies.   
In comparison, PC13 used SExtractor and auto-flux except augmented by the APEX package for objects missed by SExtractor.
This will lead to different SFR results, as shown in Sect.\ref{res}.

For the few blended sources 
(see notes on the sources in Appendix \ref{notes}), we used the \textit{aper.pro} IDL procedure\footnote{http://www.exelisvis.com/docs/aper.html} 
based one IRAF tool DAOPHOT \citep{Stetson87,Stetson92}.  For uncertainties in these cases we obtained the
uncertainty files provided by the SHA, and cross-checked our derived values 
by comparing the results we derived for unblended sources, and the results are consistent.

\subsection{Gemini spectroscopy}

We acquired three Gemini north GMOS observations (GN-2005A-Q-9, GN-2006A-Q-4, and GN-2006B-Q-38),
for which the imaging process and data reduction is described in a previous paper by \citet{Ulmer09}.
We took advantage of the
pre-imaging observation to acquire  two deep optical images of this cluster in the $i^{\prime}$ and $z^{\prime}$ bands.
We obtained images in the $i^{\prime}$ band under photometric conditions.
Images in the $z^{\prime}$ band were observed under non-photometric conditions. 

 To measure redshifts in our $z\simeq0.9$ galaxy cluster, we required spectroscopic observations
of galaxies as faint as $i^{\prime}\simeq23$. We obtained spectra of 45 galaxies with magnitudes between 
 $i^{\prime} = 20$ and $22.6$ with exposure times varying between 3 and 4 hours (within this configuration we reach 
 S/N=2.5 per resolution unit, that is $\sim$ 2.5 pixels or 7.5 \AA). 
Individual redshifts were determined through  cross-correlating the observed spectra with a set of templates of several different galaxy types. 
We used  the Ca II H\&K lines (3934\AA, 3964\AA) and [OII]  emission line (3727\AA) as the main ones for the 
redshift determination.
From the total of 45 spectra we measured, we consider cluster members  to be all  galaxies with redshift between 0.85 and 0.87 
which corresponds to a velocity dispersion of $\sim 2500~\rm km~s^{-1}$ relative to the cluster redshift (the same limits as PC13),
with uncertainties on the velocity determinations of less than $\sim$ 200 km/s.

We note that, apart from the redshifts, very little additional information on the physical properties of the galaxies can be obtained from these spectra. 
The flux calibration is very uncertain and the S/N is very low, which makes it impossible to extract meaningful information on the stellar population of 
the galaxies from these spectra.

\subsection{Near-infrared imaging}

We also obtained near infrared data ($J$ and $K_{\rm s}$ bands), which allows us to fill the gap between 
the Gemini $i^{\prime}$ and $z^{\prime}$ bands and the {\it Spitzer} IRAC and MIPS mid-infrared bands.
We obtained a $J$-band Subaru MOIRCS image with 40 min exposure in 2007. We also observed a subfield of the RX J1257 in the $K_{\rm s}$-band for two hours. 
This provided a complete catalogue up to $K_{\rm s} \simeq 22$. Details about these observations can be found in \citet{Ulmer09}.

\subsection{Cluster galaxies sample}
With the aim of building a multi-wavelength catalogue that includes all detected sources,
MIPS sources were matched with the near-IR observation catalogue using a 3$^{\prime \prime}$ matching distance, which is half of the PSF FWHM 
 at 24$\mu$m.
Overall, 78\% of objects with $i^{\prime} < 23$ and in the good spectroscopic redshift range with MIPS coverage are matched to a MIPS source. 
In cases where multiple objects in i-band could be matched with the same MIPS source, the closest object was assigned as the match. 
Of all the objects assigned to a MIPS counterpart, only two were assigned to the closest of multiple candidates.
The result is a sample of 18 galaxies spectroscopically confirmed as cluster members. The positions and spectroscopic redshifts of these 
galaxies are presented in Table~\ref{tab_res}.
We note that with our final sample of only 18 objects it is not possible to assume a rigorous completeness limit 
by searching for the turn over in the number counts vs flux. 

\section{Star Formation Rates \& Stellar Masses}
\label{sfr}

In order to investigate the star formation properties of our cluster galaxies, in this section, we derive the star formation rates of our cluster galaxies using two different methods: by fitting the full spectral energy distributions (SEDs) of our galaxies (\S\ref{magphys}), and by using the observed 24$\mu$m emission to determine the total infrared luminosity, and converting that to a SFR (\S3.2; this is a commonly used method when only infrared observations are available). 
To compute the specific SFR (sSFR), which is defined as the SFR/$M_{\star}$,
we adopt the stellar masses $M_{\star}$ derived from SED fitting for both SFR determinations.

\subsection{Determination using the MAGPHYS code}
\label{magphys}

We fit the full observed SEDs of our galaxies using the MAGPHYS code (da Cunha et al. 2008; www.iap.fr/magphys).
MAGPHYS interprets the full ultraviolet-to-infrared spectral energy distribution of galaxies in terms of several physical parameters related to the stellar content, 
star formation activity and dust content of galaxies, using a Bayesian fitting method. The stellar population is modelled using the \citet{bc03} models, assuming 
a \citet{Chabrier03} initial mass function (IMF). The dust attenuation is modelled using the two-component model of \citet{CF2000}, which includes the 
attenuation of starlight by dust in stellar birth clouds and in the diffuse interstellar medium. The energy absorbed by dust in these regions is then 
re-radiated in the mid- to far-infrared 
range via an energy balance argument, and using dust emission templates to account for different components: polycyclic aromatic hydrocarbons (PAHs), hot dust 
emitting in the mid-infrared, and warm and cold dust in thermal equilibrium \citep[more details in][]{daCunha08}. 

We use the $i^{'}$, $J$, $K_{\rm s}$, IRAC1, IRAC2, IRAC3, IRAC4 and MIPS 24$\mu$m bands as inputs for MAGPHYS.
We exclude the $z^{'}$ band because this band was observed under non-photometric conditions, and therefore the resulting fluxes are too uncertain. 
We present the resulting SED fits for each individual galaxy in Appendix \ref{magphysfigs}, and the 
the SFR and stellar masses constrained through this method in Table~\ref{tab_res}.
Error bars from MAGPHYS are the 16th--84th percentile of the likelihood distribution of SFR resulting 
from the MAGPHYS SED fit.

\subsection{Determination using the {\it Spitzer} 24$\mu$m flux}
\label{sfrmips}

In order to convert the 24$\mu$m luminosity into a total infrared luminosity, we use the technique described
in \citet{Chary01} and \citet{Elbaz07}, which is based on the observed correlation between the mid and far infrared luminosity
of local galaxies. A library of 105 template spectral energy distribution (SEDs) was built to reproduce those correlations
including the tight correlation between the rest-frame $\sim$13$\mu$m luminosity 
with $L_{\rm IR}$, probed by the observed 24$\mu$m 
passband for galaxies at $z\simeq0.9$. For each cluster galaxy, we compute the rest-frame luminosity at $24 \mu m/(1+z)$ and compare it
to the luminosity at that wavelength for each one of the 105 template SEDs, which allows us to identify the template with the closest luminosity. 
We then use this template, normalised to the observed luminosity of the galaxy at $24 \mu m/(1+z)$, to derive $L_{\rm IR}$ for that specific galaxy.
To do so, we use the \citet{kennicutt98} relation: $\rm SFR_{\rm IR} [M_{\odot} \, yr^{-1}] = 1.72 \times 10^{-10}\, L_{\rm IR} [L_{\odot}]$.

We note that different IMFs lead to systematically different SFR derivations.
The MAGPHYS SFR determination, described in the previous section, uses a Chabrier IMF, while the \citet{kennicutt98} relation
assumes a Salpeter IMF \citep{Salpeter55}. 
Thus, in order to make these two different determinations comparable, we rescale the SFR computed from {\it Spitzer} observations,
using SFR[Salpeter] =  $1.7\times$ SFR[Chabrier] \citep[e.g.,][]{Genzel10}.
The resulting SFR and sSFR for each galaxy obtained using this method are also presented in Table~\ref{tab_res}.

\section{Results}
\label{res}

\subsection{Comparison of different SFR estimates}

In this section we compare the SFRs of our cluster galaxies obtained using the two methods described in the previous section.

\begin{figure}[ht!]
 \centering
 \includegraphics[width=0.35\textwidth]{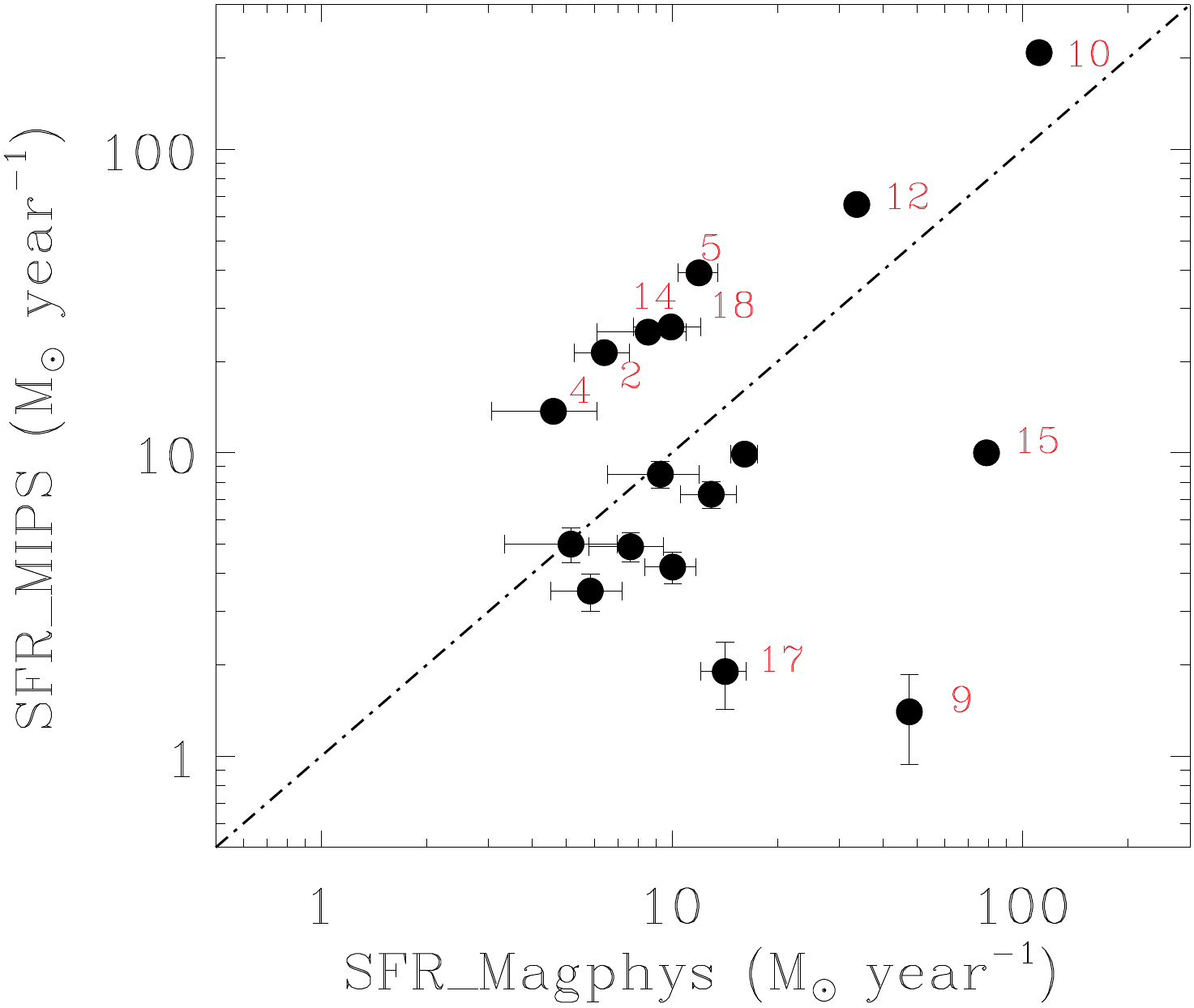} 
 \caption{Comparison of the SFR values computed from the total IR luminosity
extrapolated using the MIPS flux (\S3.2) and the MAGPHYS code output (\S\ref{magphys}). In red we label the IDs of the sources with
the largest differences between the two SFR estimates.The dashed line is the identity line.}
  \label{fig_mipsmagphys}
\end{figure}

We note that the MIPS-based and MAGPHYS estimates of the SFR are based on fundamentally different assumptions.
The MIPS 24$\mu$m flux traces the rest-frame mid-infrared (around $13\mu$m) of our cluster galaxies, which is dominated
by a combination of hot dust and PAHs emission features, both heated by ultraviolet radiation emitted by young stars (i.e. 
ongoing star formation). However, there are caveats in using this wavelength regime to constrain the SFR, 
specifically the fact that PAH emission depends strongly on metallicity, and that these dust components
can also be heated by evolved stellar populations unrelated to current star formation \citep{calzetti10}.
The MIPS-based SFR estimates also suffer from the uncertainty in extrapolating from the rest-frame mid-infrared to
the total infrared emission, which assumes that galaxy infrared SEDs do not vary significantly out to $z\simeq1$.

The MAGPHYS modelling, on the other hand, interprets the full ultraviolet-to-infrared spectral energy distribution 
of galaxies in a self-consistent way, taking into account dust heating by
young stars mostly associated with stellar birth clouds, and also heating by
older stellar populations in the diffuse ISM. The extrapolation of the dust SEDs to wavelengths longer than 24$\mu$m is done
using a physically-motivated energy balance technique and assuming a large variety of dust emission properties (such as temperatures,
PAH fractions, etc.).
In some cases, the presence of an AGN may also provide additional mid-infrared emission (from the heating of the dusty torus)
which can contaminate the SFR estimates if not taken into account. This affects both the MIPS-based and MAGPHYS estimates of the SFR.
Our SEDs (see Figs.\ref{fig_magphys}-\ref{fig_magphys2}) point to little or no AGN contamination, when compared
to Fig.1 from \citet{Donley12}. Also, when taking the IRAC fluxes of these galaxies and putting them in an AGN mid-IR diagnostic plot 
\citep[e.g.][but based on Fig.4 from the later reference]{Stern05,Lacy07,Donley12} none of them lie in the current  AGN selection regions.

In Fig.\ref{fig_mipsmagphys}, we show the comparison between the SFR derived from MIPS observation and using the MAGPHYS code.
Galaxies with discrepant SFR values between the methods are identified in this figure using the source codes
from Tab.\ref{tab_res}.
The SFR obtained from MIPS is significantly lower than the one obtained from MAGPHYS for galaxies identified 
as 9, 15 and 17.
The reason they fall so far below the locus of the other galaxies may come from the different inferred total infrared luminosities.
For these three galaxies the $L_{\rm IR}$ derived from MAGPHYS is about one order of magnitude higher than the one computed using
the MIPS flux, which could be due to the energy balance based on the stellar population modelling implying much higher dust
luminosities than the dust SEDs \citet{Chary01} templates.

We can also speculate that this could be a metallicity effect such that PAHs can be weak (as shown by the MAGPHYS fits in Figs.\ref{fig_magphys}-\ref{fig_magphys2})
 or even absent in low metallicity star forming galaxies \citep{Engelbracht05,Madden06}, or due to PAH destruction \citep[e.g.,][]{Giard94}.

On the other hand, for many galaxies (identified as 2, 4, 5, 10, 12, 14 and 18) the opposite behaviour is seen, where the SFR from MIPS 
is higher than the one obtained from MAGPHYS. 
One possible explanation is that these are galaxies in crowded regions, and most of them have at least one very close companion, 
whose MIPS flux might be contaminating the SFR determination (which is the case for galaxies identified as 2, 10, 12, 14, and 18).
This explanation cannot be applied for two of these galaxies, identified as 
4 and 5, however. Another equally likely explanation could be again related to the uncertainties
on the relation of the $L_{\rm IR}$ with the SFR, but comparing the inferred total infrared luminosities,
for galaxy identified as 4, the $L_{\rm IR}$ obtained from MAGPHYS is about 2 times higher than the $L_{\rm IR}$ derived from MIPS flux
and for galaxy identified as 5, the $L_{\rm IR}$ computed with these two different methods agree. 
Then, the difference in SFR is probably due to different assumptions about dust heating. The IR method assumes that all of the dust emission is heated
by current SF (i.e. young stars), but MAGPHYS allows for some of the dust emission to be heated by older stellar populations therefore this
allows for lower SFR.

\subsection{Dependence of star formation rates on environment and stellar mass}
\label{sfr2Rmstar}

\begin{figure*}[ht!]
 \centering
 \includegraphics[width=0.75\textwidth]{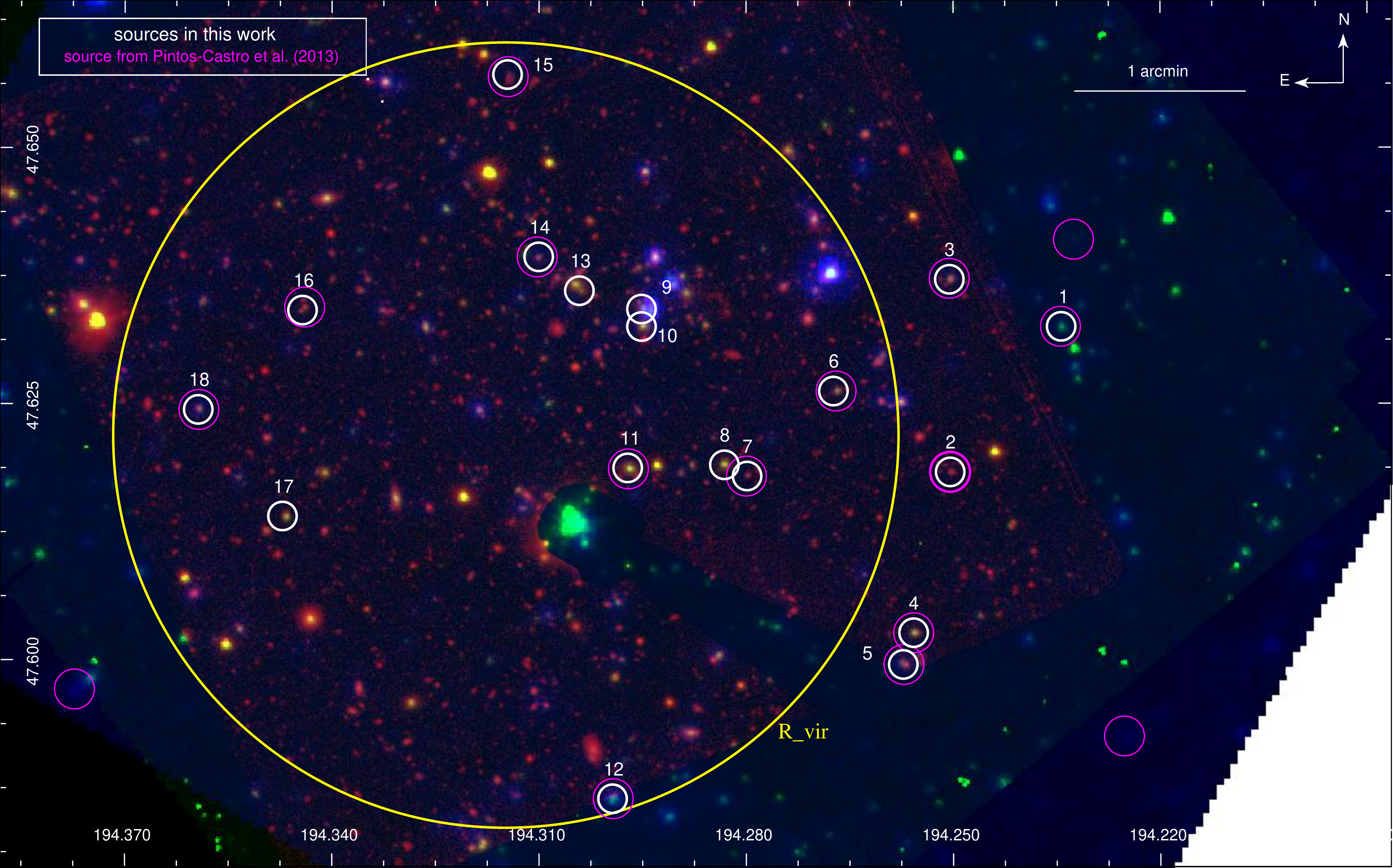} 
 \caption{Color image of our cluster based on Gemini i'band image (R),  {\it Spitzer} {\color{blue} IRAC1} (G) and {\it Spitzer} MIPS {\color{blue} 24$\mu$m} (B). 
 White circles represent the galaxies within the cluster redshift range used in this work, 
and the large yellow circle correspond to the virial radius of 1.05 Mpc.
 The magenta circles represent the galaxies at the cluster redshift range in \citet{PintosCastro13}. 
 We have 11 out of 18 galaxies in common.}
  \label{fig_regs}
\end{figure*}

In this section, we investigate in detail how the star formation activity of our cluster galaxies depends on their environment by looking at their position
within the cluster.
In Fig.~\ref{fig_regs}, we present the spatial distribution of the galaxies used in this work and a comparison with the ones used
in PC13. We note that PC13 considered 3 galaxies with no MIPS detection: R.A.$=194.2251$, Dec$=47.5925$;
R.A.$=194.2324$, Dec$=47.641$, and R.A.$=194.3771$, Dec$=47.5971$ (the open magenta circles with no corresponding white circle in Fig.\ref{fig_regs}), and
they missed five galaxies in the innermost part of the cluster (inside $r_{\rm vir}$).  
Out of these, two were not identified by PC13 because they are a little more than 1 mag below their completeness limit (objects 9 and 13 in Table \ref{tab_res}).  
Object 10 however, was possibly rejected by PC13 not because it is too faint, but because they associated its flux to a line-of-sight, non-cluster member.  
Other objects present in our sample that cannot be found in PC13 are objects 8 and 17.  Object 8 (mag 19.6) is just below the PC13 completeness 
limit and object 17 (mag 20.2) is well below. 
Thus, only object 10 was not picked up by PC13 either because of image artifacts or because they attributed the MIPS flux to another galaxy. 
The different completeness limit suggests that the data used by PC13 was not the latest data set processed through the 
\textit{Spitzer} pipeline, as mentioned before. Despite that, we still have 11 sources in common with their work.

\begin{figure}[ht!]
 \centering
 \includegraphics[width=0.35\textwidth,angle=90]{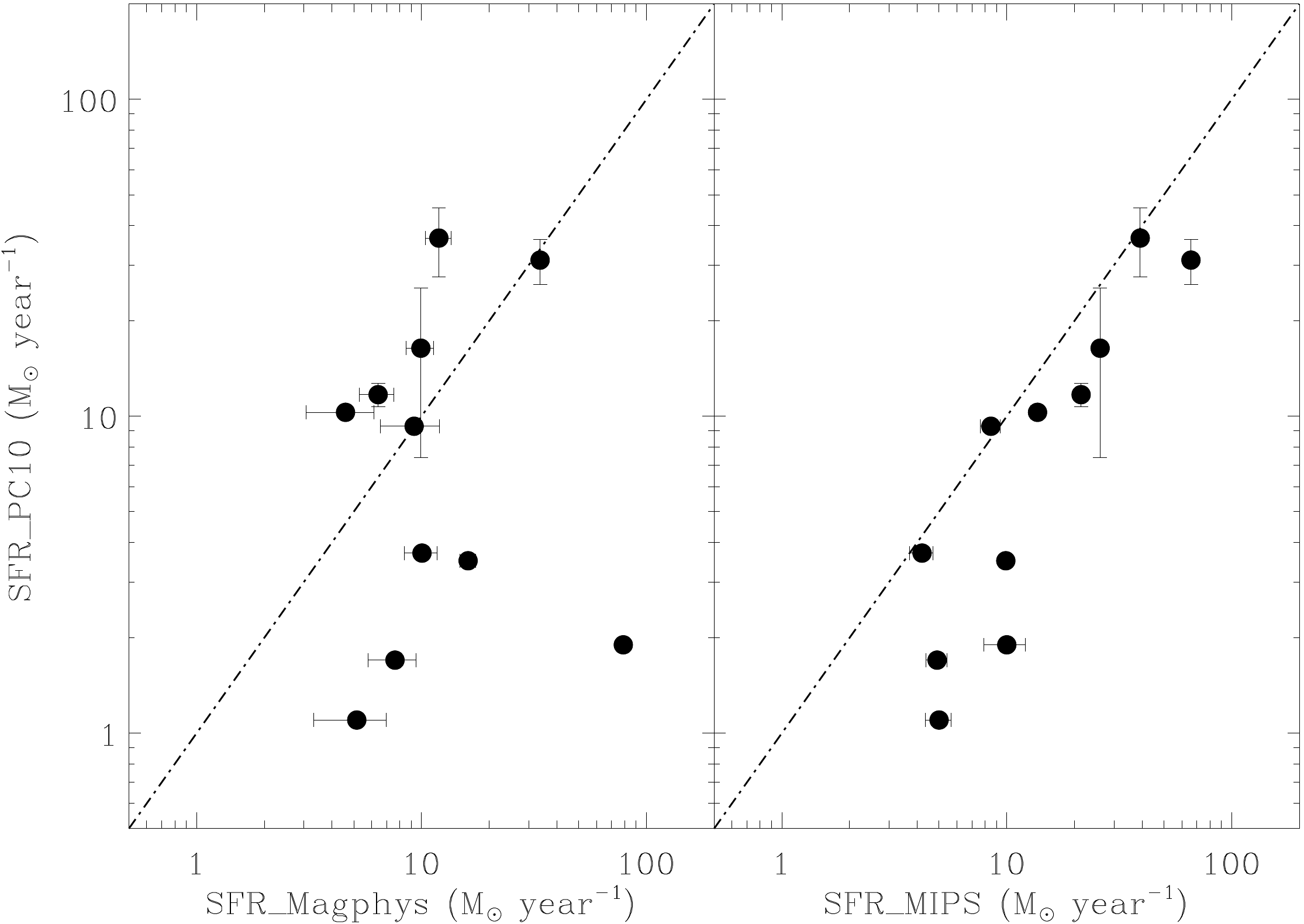} 
 \caption{Comparison of the 11 sources we have in common with \citet{PintosCastro13},  we present a comparison of their SFR values against
 the one presented here determined using  MAGPHYS code (left panel) and MIPS (right panel). The dot-dashed lines represent the 
 one-to-one correspondence.}
  \label{fig_sfrcomp}
\end{figure}

For the 11 sources in common, Fig.\ref{fig_sfrcomp} shows the comparison between the SFR values derived here 
and the ones obtained by PC13. 
As expected, PC13 SFR results agree better with the MIPS SFR, since they also obtained
their SFR from the total $L_{\rm IR}$, which in turn was obtained by fitting the same SED templates
used here, although we can notice some systematic offset of PC13 SFR values being lower than MIPS
SFR, specially for SFR $< 5 M_{\odot}$ sun/yr. As mentioned before, PC13 used a shallower dataset 
and the methods to measure the flux density were different. Thus, the faintest sources are the 
ones with greater differences in SFR determination between our method and PC13.

\begin{table*}
\footnotesize
\begin{center}
\caption{Properties of the cluster galaxies.\label{tab_res}}
\begin{tabular}{cccccccccc}
\hline
\hline
Source  & redshift & RA & DEC &  $\rm SFR_{\rm MAGPHYS}$  & $M_{\rm star_{MAGPHYS}}$ & $\rm flux_{\rm MIPS} $  &     $\rm SFR_{\rm MIPS}$ &  sSFR   & projected dist. \\     
 code & & (J2000) &  (J2000) & ($\rm M_{\odot}/yr$) & ($10^{11} M_{\odot}$)& ($\mu$ Jy)  &($M_{\odot}/yr$) &  ($10^{-10} yr^{-1}$)  & (Mpc)   \\
\hline
\end{tabular}
\begin{tabular}{lccc  r@{.}l r@{.}l  r@{.}l  r@{.}l  c c}
1	&	0.874	&	194.2342	&	47.6325 & 9&$3^{+2.6}_{-2.9}$    & 1&15$^{+0.39}_{-0.47}$   & 98&8	 $\pm$	10.0	&		8&5$\pm$0.9	&	0.86 $\pm$ 0.27&	1.26 \\
2	&	0.865	&	194.2500	&	47.6371 &  6&$4^{+1.1}_{-1.1}$   & 0&98$^{+0.16}_{-0.15}$  &  218&7	$\pm$	6.9	&		21&4$\pm$0.7	&	3.26	$\pm$ 0.48 &	0.97 \\
3	&	0.871	&	194.2503	&	47.6183 & 5&$2^{+2.0}_{-1.2}$    & 0&15$^{+0.02}_{-0.03}$  & 64&8	$\pm$	8.3	&		5&0$\pm$0.6	&	3.11	$\pm$  0.96&	1.05 \\
4	&	0.853	&	194.2556	&	47.6026 & 4&$6^{+1.6}_{-1.4}$    & 1&70$^{+0.29}_{-0.25}$  & 151&0	$\pm$	10.1	&		13&7$\pm$0.9	&	1.41	$\pm$ 0.39 &	0.96  \\
5	&	0.853	&	194.2571	&	47.5995 & 11&$9 ^{+1.5}_{-1.6}$ & 1&07$^{+0.22}_{-0.28}$  & 383&2 	$\pm$	10.3	&		39&2$\pm$1.1	&	6.30	$\pm$ 1.25 &	1.22   \\
6	&	0.863	&	194.2670	&	47.6262 & 10&$0^{+1.7}_{-1.7}$  & 0&85$^{+0.11}_{-0.15}$  & 56&2 	$\pm$	6.9	&		4&2$\pm$0.5	&	0.76	$\pm$ 0.20 &	1.27   \\
7	&	0.873	&	194.2797	&	47.6179 & 7&$6^{+1.6}_{-2.0}$    & 0&07$^{+0.02}_{-0.02}$  & 63&8      $\pm$	7.0	&		4&9$\pm$0.5	&	3.66	$\pm$ 0.95 &	0.60  \\
8	&	0.858	&	194.2830	&	47.6190 & 5&$8^{+1.3}_{-1.3}$    & 2&04$^{+0.30}_{-0.20}$   & 48&4	$\pm$	7.0	&		3&5$\pm$0.5	&	0.27	$\pm$ 0.07 &	0.54  \\
9	&	0.862	&	194.2950	&	47.6325 & 47&$5^{+1.4}_{-1.4}$  & 2&51$^{+0.52}_{-0.31}$  & 20&7	$\pm$	6.9	&		1&4$\pm$0.5	&	0.09	$\pm$ 0.04 &	0.13   \\
10	&	0.864	&	194.2950	&	47.6342 & 111&$4^{+2.3}_{-2.1}$& 6&17$^{+0.28}_{-0.29}$  & 1642&1 	$\pm$	6.9	&		208&7$\pm$0.9 &	13.11 $\pm$ 2.01 &	0.12   \\
11	&	0.856	&	194.2970	&	47.6187& 16&$1^{+1.4}_{-1.4}$   & 2&45$^{+0.37}_{-0.30}$  & 109&8	$\pm$	7.0	&		9&9$\pm$	0.6   &	0.55	$\pm$ 0.11 &	0.43    \\
12	&	0.865	&	194.2992	&	47.5864 &  33&$7^{+1.6}_{-1.7}$ & 1&23$^{+0.42}_{-0.51}$  & 591&8	$\pm$	10.0	&		65&9$\pm$1.1   &     5.31	$\pm$ 0.91 &	1.32   \\
13	&	0.865	&	194.3040	&	47.6360 & 	- & -			     & 		-&-		                & 21&0	$\pm$	6.9	&		1&4$\pm$0.5	&	0.07	$\pm$ 0.03 &	0.07  \\
14	&	0.853	&	194.3099	&	47.6393&  8&$6^{+2.4}_{-2.5}$     &1&00$^{+0.38}_{-0.48}$  &  259&7	$\pm$	7.1	&		25&0$\pm$0.7	&	7.19	$\pm$ 1.14 &	0.22  \\
15	&	0.860	&	194.3142	&	47.6571&  78&$9^{+1.5}_{-1.0}$   & 0&11$^{+0.05}_{-0.01}$ & 112&0	$\pm$	7.6	&		10&0$\pm$0.7   &	4.07	$\pm$ 0.96 &	0.69   \\
16	&	0.860	&	194.3441	&	47.6341&  12&$9^{+1.8}_{-2.9}$   & 0&31$^{+0.07}_{-0.09}$ & 92&4 	$\pm$	9.5	&		7&3$\pm$0.7	&	2.22	$\pm$ 0.64 &	0.80  \\
17	&	0.859	&	194.3470	&	47.6140 & 14&$2^{+2.3}_{-1.9}$   & 1&29$^{+0.19}_{-0.22}$ & 28&1	$\pm$	6.9	&		1&9$\pm$0.5	&	0.22	$\pm$ 0.08 &	1.02  \\
18	&	0.860	&	194.3592	&	47.6244 &  9&$9^{+2.2}_{-2.11}$  & 0&72$^{+0.30}_{-0.42}$ & 264&2	$\pm$	13.5	&		26&0$\pm$1.3	&	 6.26	$\pm$ 1.70 &	1.11   \\
\hline
\hline
\end{tabular}
\end{center}
\end{table*}

\begin{figure*}[ht!]
 \centering
 \includegraphics[width=0.35\textwidth,angle=90]{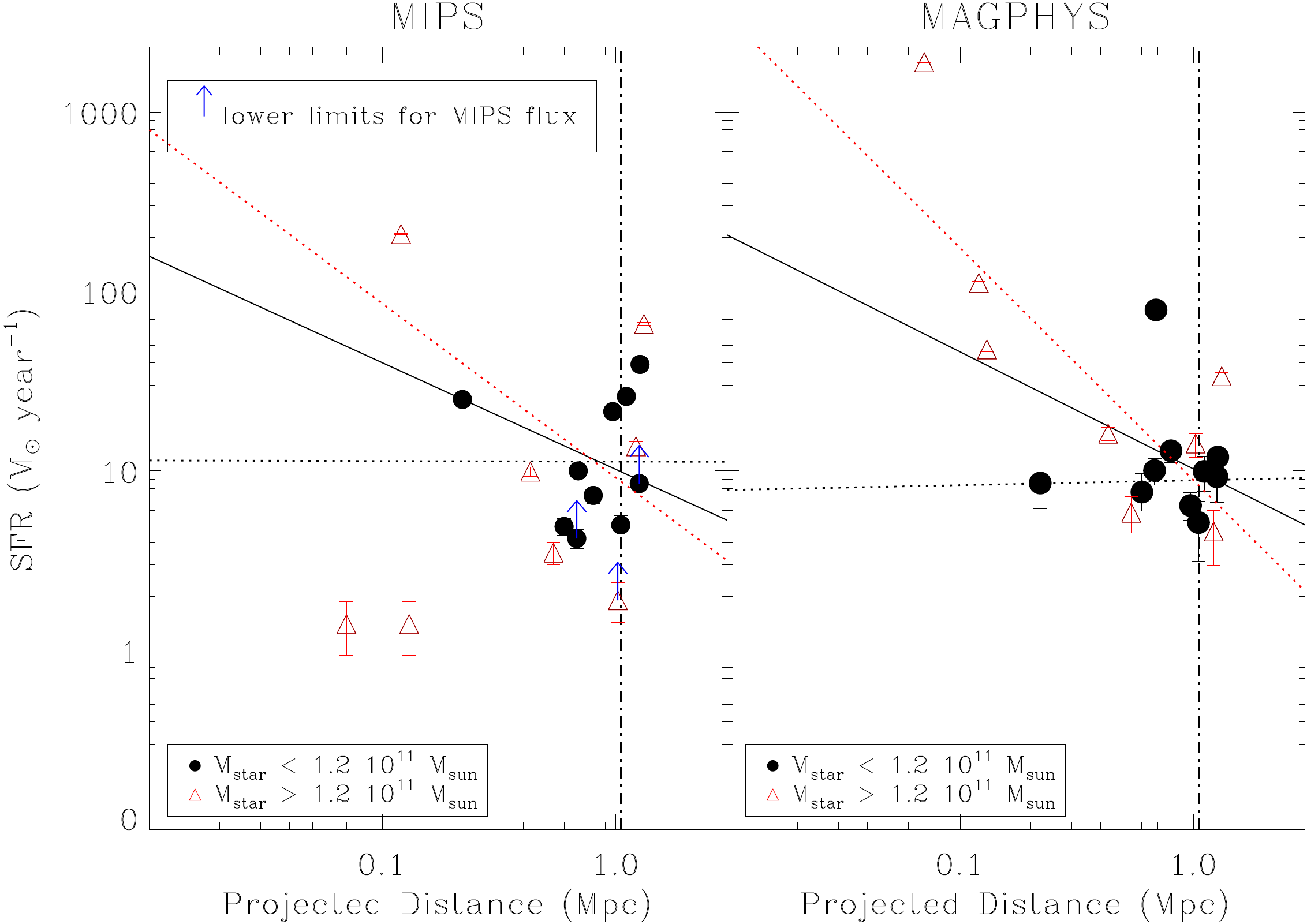} 
  \includegraphics[width=0.35\textwidth,angle=90]{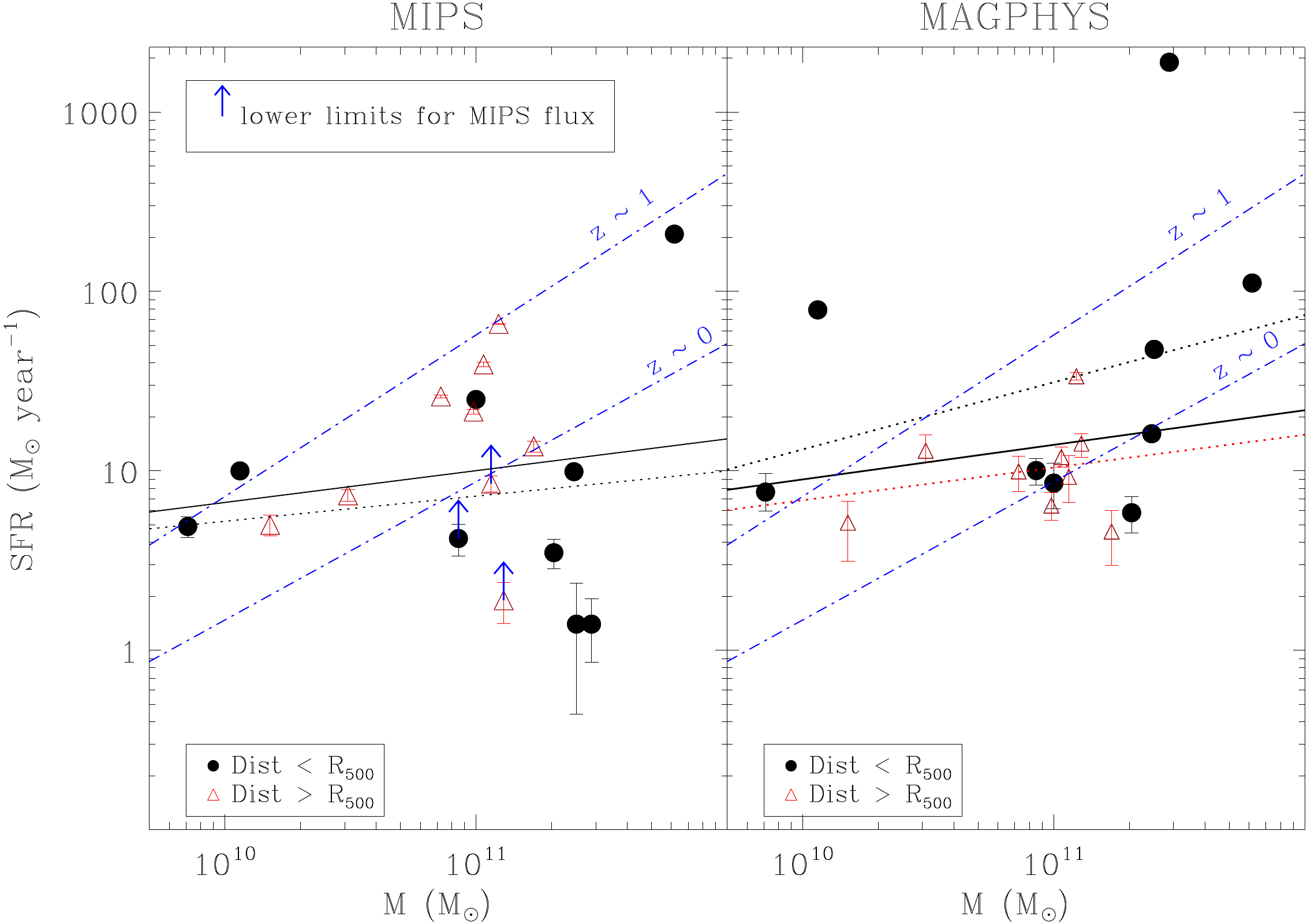} 
 \caption{Star formation rates as a function of the projected distance from the center (left panels) and stellar masses (right panels),
 for SFRs computed from MIPS and using MAGPHYS code. 
 On panels in which the SFR was computed from MIPS flux, we overplot blue upward-arrows for the sources we used reduced apertures to get the flux due 
 to nearby sources).  
 The solid black lines correspond to the best fit for the whole sample. The red-dotted lines represent the best fit for the red points and the
 black-dotted lines are the best-fit for the black points.
 On the left panels, the black-filed circles represent less massive galaxies  with stellar masses $M_{\star} < 1.2 \times 10^{11} M_{\odot}$, 
 the open-red triangles  represent more massive galaxies,  with $M_{\star} > 1.2 \times 10^{11} M_{\odot}$. 
 On the right panels, the black-filed circles represent the galaxies within $r_{500}$ ($\sim$ 2/3 viral radius) and the open-red triangles represent  galaxies
 outside $r_{500}$. The  dot-dashed blue lines correspond to the field trend at $z=0,1$, respectively.}
  \label{fig_sfr}
\end{figure*}

We now analyse how the SFR and the specific SFR of our galaxies depends on environment.
Here, it is important to highlight that we often find in the literature two different proxies for environment: the cluster centric radius and the local galaxy density, 
which is calculated from the distance to the 10th nearest neighbour \citep[e.g.,][]{Noble13}. The former, which is used in this work, is a better indicator of the global 
environment and cluster potential.  Although the local density has a proximate effect on galaxies, the cluster has been in existence about 2.5 Gyr. 
Thus, the current location may not be a good indicator of the galaxy's average local environment.
We note that our radial measurements do not require a completeness correction since
 the completeness bias for mass is similar at all radii.
 
To analyse our correlations, we applied a robust linear regression fit and a Pearson linear correlation coefficient, which we list in Appendix~\ref{tab_fit}.
In the left panel of Fig. \ref{fig_sfr}, we plot SFR of each of our cluster galaxies as a function of its projected cluster centric distance (our proxy for
 the cluster environment), for the SFR derived from MIPS observations (left-hand panel) and using the MAGPHYS code (right-hand panel). 
 The black-solid lines represent the best fit for the whole sample.
 The different symbols represent galaxies of different stellar masses. Black-solid points represent less massive galaxies with $M_{\star} < 1.2 \times 10^{11} M_{\odot}$ 
 (the median value of our sample) and the black dotted line represents the best fit for these galaxies. 
 The red-open triangles represent the most massive galaxies with masses 
 $M_{\star} \geq 1.2 \times 10^{11} M_{\odot}$ and,
 the red dotted line represents the best fit for these more massive galaxies. 
 The plot that correspond to MAGPHYS SFRs shows that the SFR decreases with projected distance, in which galaxies at the center having higher SFR. 
 For the $\rm SFR_{\rm MIPS}$-Distance relation, if we look 
 at distances larger than 0.2 Mpc, where most galaxies are, it is hard to see a dependence of the SFR on the projected distance.
 When we separate galaxies by mass, 
 the SFR-Distance relation is steeper for more massive galaxies (with a linear Pearson correlation coefficient of -0.78) than for the less
 massive ones, in which  SFR-Distance is consistent with no correlation. Here, we reinforce the fact that for higher mass galaxies the two innermost
 galaxies are driving the correlation. If we consider galaxies at distances larger than 0.2 Mpc, there is no correlation between SFR and Distance.
 
In the right panel of Fig. \ref{fig_sfr}, we plot the SFR as a function of the galaxy stellar mass for our galaxies.
 The black solid lines represent the best fit for the sample. The different symbols represent galaxies in different radii, where black-solid points represent 
 galaxies inside $r_{500}$ and the black dotted line represents the best fit for these galaxies. 
 The open-red triangles represent galaxies outside $r_{500}$ and, the red dotted line represents the best fit for these galaxies. 
 For these panels we see that the  SFR-$M_{\star}$ relation presents a mild increase (with a Pearson correlation coefficient of about 0.20
 for both methods). 
 However, when we separate galaxies by
 distance, the SFR presents the same dependence on stellar mass for galaxies inside and outside $r_{500}$ and both
 being consistent with no correlation.
 This indicates that the galaxies maintain almost the same star formation at all densities
 \citep[consistent with previous studies such as][]{Peng10,Lu12,Muzzin12}.

We also show as a comparison two blue dot-dashed lines in the right panels of Fig.\ref{fig_sfr} that correspond to the main-sequence trend in 
  the field at two different redshifts:
 0.1 from the Sloan Digital Sky Survey \citep[SDSS;][]{Brinchmann04} as analysed by \citet{Elbaz07} and 
 the blue star forming galaxies at $z=0.8-1.2$ from The Great Observatories Origins Deep Survey \citep[GOODS;][]{Elbaz07}. 
These lines are important to guide our analysis. What can be seen from this figure is that the SFR of 
most galaxies, is closer to the main-sequence trend observed in the z $\sim$ 0 SDSS field than to the main-sequence 
trend observed in the z $\sim$ 1 GOODS field galaxies (although for MIPS SFRs there are $\sim 1/3$ close to z $\sim$ 1).
For the MAGPHYS SFRs, the few galaxies closer to the  $z \sim 1$ trend,  are the less massive ones and there seems to 
be no preference of location in the cluster.

\begin{figure*}[ht!]
 \centering
  \includegraphics[width=0.36\textwidth,angle=90]{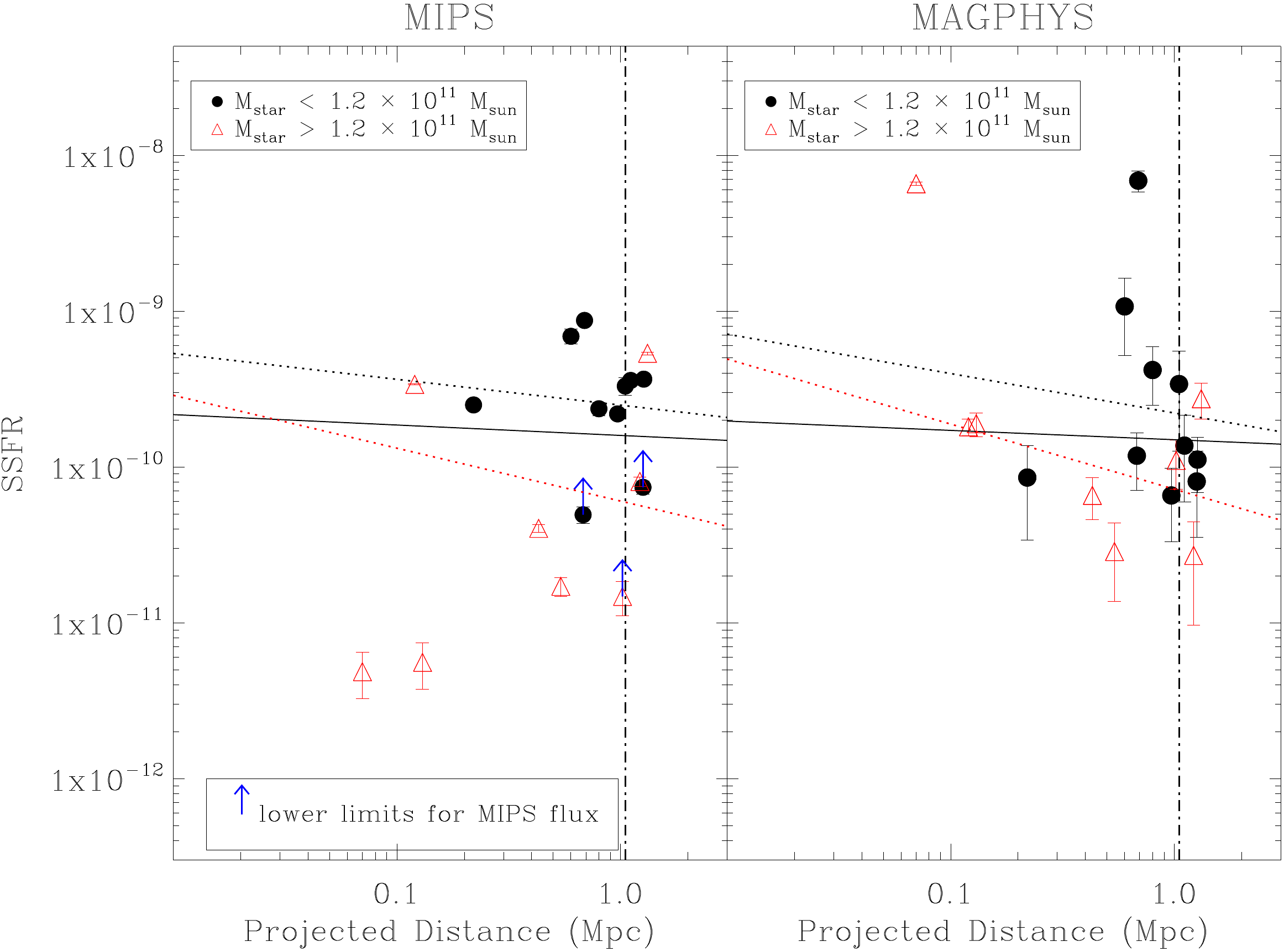} 
 \includegraphics[width=0.36\textwidth,angle=90]{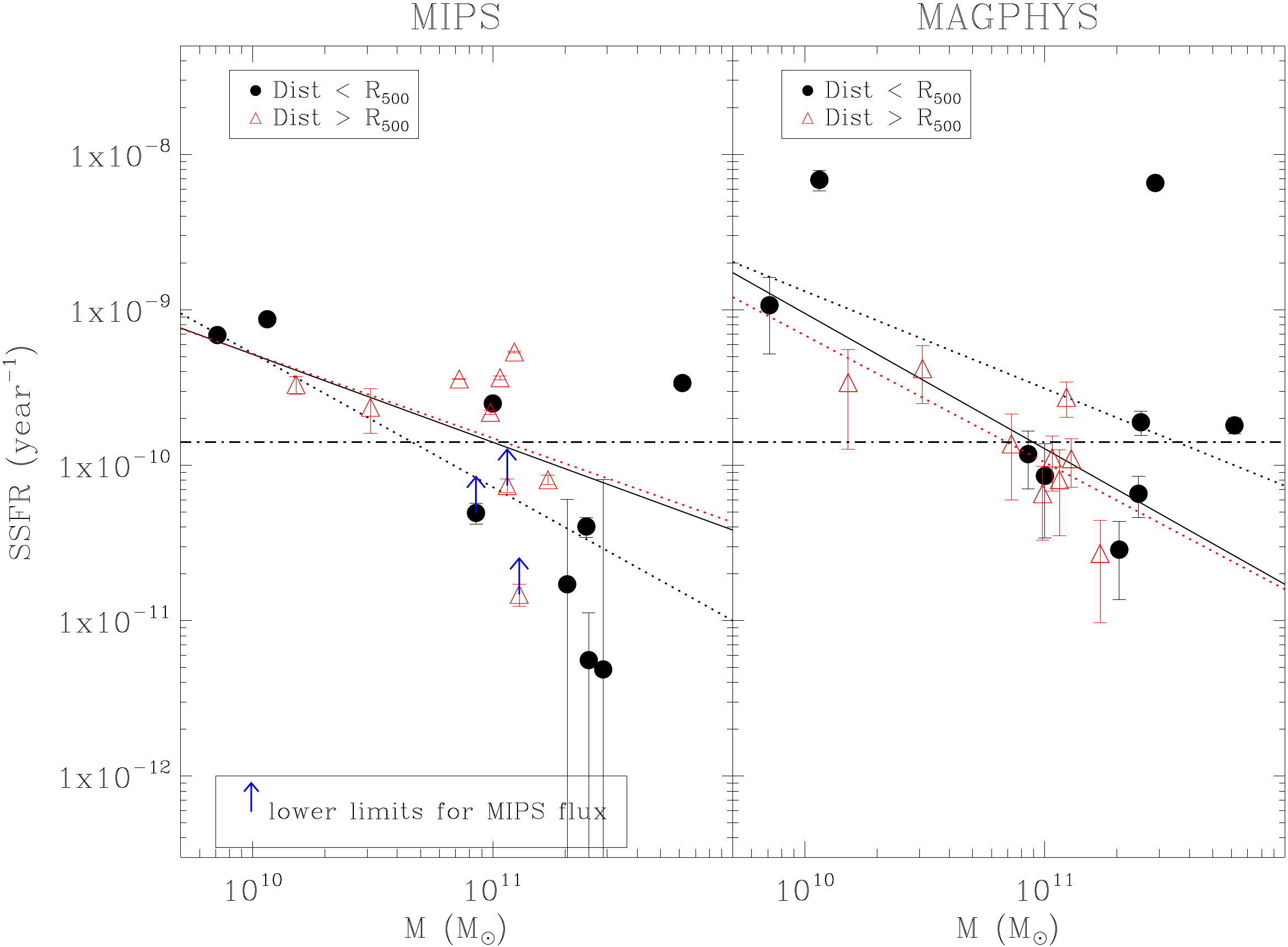} 
  \caption{Specific star formation rates as a function of projected distance from the center (left panel)  and stellar mass (right penal). 
The symbols follow the ones in Fig.\ref{fig_sfr}. The dot-dashed horizontal line in te right panels represents the sSFR 
 at which stellar mass doubles by at $z=0$ if the SFR remains the same \citep{Patel09}.}
  \label{fig_sSFR}
\end{figure*}

In principle, Fig.\ref{fig_sfr} suggests that SFR is not related to the environment (traced by the cluster centric distance) but to galaxy mass.
But when  trying to understand the role of environment on galaxy activity, one 
has to be aware that 
 the most massive galaxies are located preferentially in denser environments. 
It is therefore difficult to separate the effect of external conditions 
from the increased in-situ activity due to a larger galaxy mass. 
Thus, in hopes of identifying if indeed more massive galaxies have enhanced SFR, 
we have plotted in Fig.\ref{fig_sSFR}, the specific SFR (SSFR) 
against the projected cluster centric distance and stellar mass.

As we can see from the left panel of Fig.\ref{fig_sSFR} there is no dependence of the SSFR with cluster distance,
but  the SSFR is  decreasing with stellar
mass (Pearson correlation coefficient of -0.75), meaning that more massive galaxies are forming less stars per unit of mass,
and most likely, these are the galaxies  that will populate the red-sequence of clusters.
This is consistent with the evolution of the mass function \citep[e.g.,][]{Kodama04} which shows that the massive end of galaxy mass function in cluster
is in place by z $\sim$1 and the evolution between z=0-1 consists of a buildup of the $\le 10^{11} M_{\odot}$ end.
We also show in this figure that the bulk of SF has not yet taken place for approximately  half of the galaxies, 
since most of them lie above the dot-dashed line that indicates the SSFR at which the stellar mass doubles by at $z=0$ if the SFR remains constant
\citep{Patel09}.

\section{Discussion}
\label{disc}
We have analyzed a single z=0.866 young galaxy cluster from multi-wavelength perspective, and we investigated the dependence of star formation activity as a function of 
cluster centric distance (a proxy for environment density) and stellar mass for 18 spectroscopically confirmed members.
We computed the star formation rates using two independent methods: extrapolating the observed 24$\mu$m emission to get the total infrared luminosity, and modelling the full SEDs with the MAGPHYS code.


Our work allows us to assess the role of the environment and stellar mass on star-formation rates and we find that MIPS cluster members share certain
properties with field galaxies, presenting a SFR that follows $z \sim 0$ field galaxies, and most of them populate the outskirts of the cluster.
In fact, a non negligible fraction of cluster members are forming stars actively.

\begin{figure}[ht!]
 \centering
  \includegraphics[width=0.3\textwidth,angle=90]{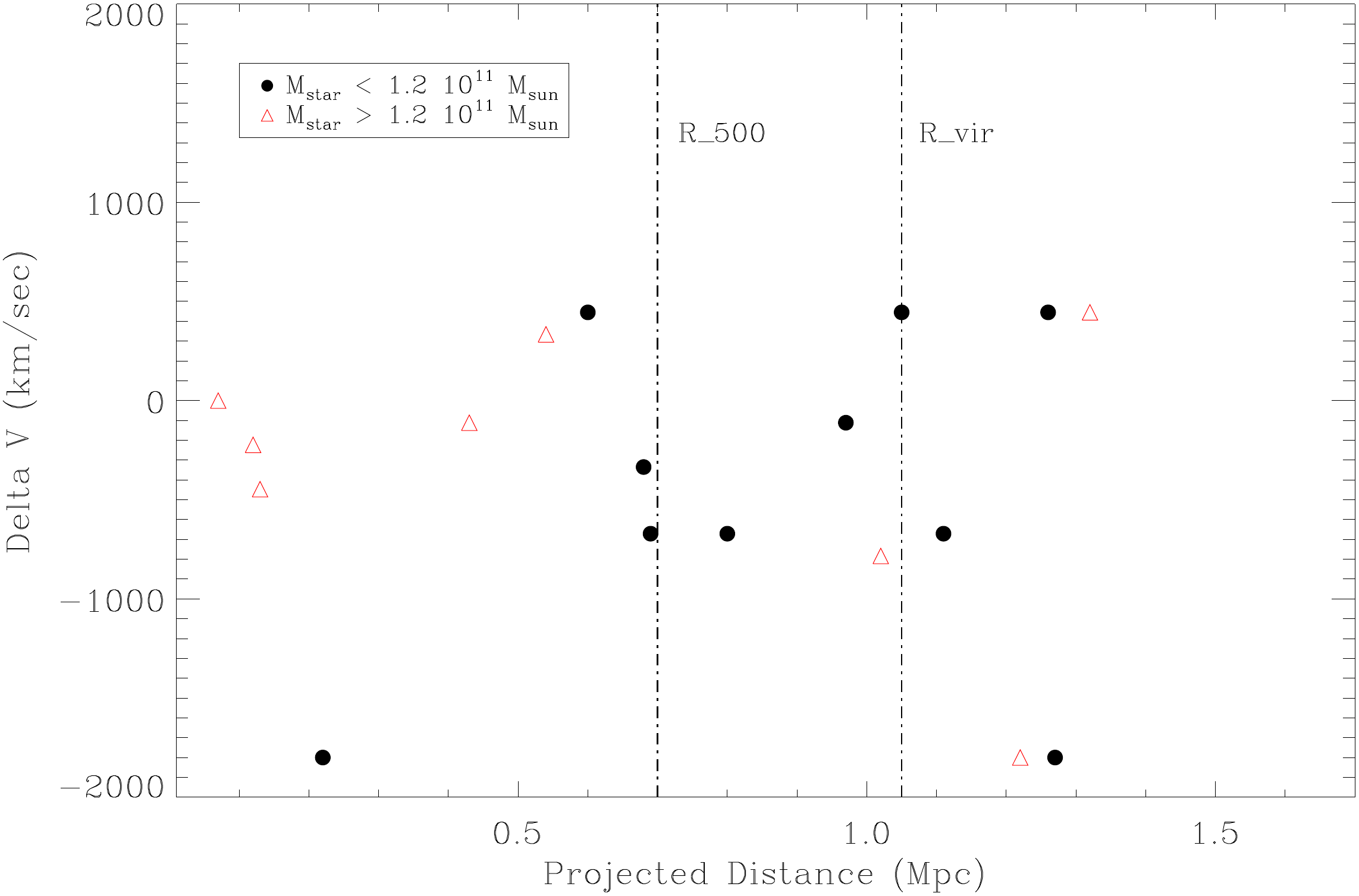} 
  \caption{Absolute line-of-sight velocity as a function of projected radius. The two vertical dot-dashed lines correspond to $r_{500}$ and $r_{\rm vir}$ and the different symbols represent galaxies
  of different mass with color identification as the left panels of Fig.\ref{fig_sfr} and Fig.\ref{fig_sSFR}.}
  \label{fig_VelDist}
\end{figure}

If the SFRs are higher for more massive galaxies, when we investigate the SSFRs as a function of stellar mass, 
we find out that it is the less massive galaxies that are forming the bulk of stars. 
If the environment is somehow driving the SF, one would expect a relation between the SSFR and the
cluster centric distance, but that is not the case. This result offers a hint alternative interpretation, as suggested by \citet{Noble13}: 
it indicates that inner regions of the cluster are not only populated by galaxies that has already spent a great time in in dense environments, but it 
is  also polluted by infalling high-velocity galaxies that could be on their pass through the
cluster center and therefore not virialized with the core population.
This can be seen by the black filled-circles in Fig.\ref{fig_VelDist}
that present velocity dispersion of about 600 km/s (and a galaxy near the cluster center with absolute line-of-sight velocity of almost of 2000 km/s). 
As these systems have retained a nominal level of star formation for their stellar mass, they enhance the average 
SSFR in the inner radial bin (cluster core).

As already  reported in \citet{Ulmer09}, this cluster is currently in the process of formation, presenting a bimodal distribution of X-ray emission, the presence of substructures 
and a high $kT$ compared to the $\rm L_{\rm X, bol} - kT$ relation. This process of build-up can explain most of the results found here.
A non negligible fraction of cluster members are forming stars actively, and that infall is probably responsible for much of the SF activity we see in the cluster.
It is unlikely that cluster only passively accretes star-forming galaxies
from the surrounding field, and those galaxies have a higher level of SF simply because they were field galaxies. On the contrary, 
since we analysed these relations as a function of the cluster centric distance that is a tracer of the global environment and potential well, 
it is possible that we are seeing an increased SFR in infalling galaxies triggered by the galaxy-IGM interaction as shown by previous theoretical and observational evidence 
\citep[e.g.,][]{Gavazzi95,Bai07}.

\section{Conclusions}
\label{conc}

 Our work highlights the significant role of the environment in the star formation activity of the young cluster RX J1257+4738.
 We use the deepest existing mid-infrared survey at 24$\mu m$, with some of the largest spectroscopic 
 completeness to analyse the relationship between the SFR and sSFR as a function of projected cluster centric distance 
 (as a proxy of the density) in a galaxy cluster at z$\sim$0.9. In summary, our analysis lead to the following conclusions:

\begin{itemize}

\item{We showed that when we separate galaxies by mass, more massive galaxies tend to have a higher SFR than less massive ones. 
However, the SFR-$M_{\star}$ relation is independent of the projected cluster centric radius, indicating that galaxies display almost the 
same SF at all densities. In principle these results indicate that SFR is not related to environment but to stellar mass.}
 
\item{In order to understand whether the SFR-density relation is generated by the presence of the massive galaxies in denser regions, we showed that
the SSFR slowly decreases with stellar mass, showing that the less massive galaxies are forming the bulk of stars.
This indicates that the increase in star-formation is driven by cluster assembly and galaxy infall. If environment is driving star formation, one would expect a relation between
the SSFR and the projected distance, that is not observed in our data.}

\item{A possible scenario to explain this lack of correlation is the contamination by high-velocity galaxies in the inner part of the cluster which could be 
infalling galaxies that may be on their initial pass through the cluster center and therefore are still not virialized with the core population. As these systems have retained
a nominal level of star-formation for their stellar mass, they enhance the mean SSFR in the center (inside $r_{500}$).}



\end{itemize}
 
Partially based on observations obtained at the Gemini
Observatory which is operated by the Association of Universities for Research in Astronomy, Inc., under cooperative agreement with the NSF on behalf 
of the Gemini partnership: the National Science Foundation (United States), the Science and Technology Facilities Council (united Kingdom), the National Research
Council (Canada), CONICYT (Chile), the Australian Research Council (Australia), Minist\'{e}rio da Ci\^{e}nice e Tecnologia (Brazil) and SECYT (Argentina). 
Partially based on observations made with the \textit{Spitzer} Space Telescope, which is operated by the Jet Propulsion Laboratory, California Institute of 
Technology under contract with NASA.

\begin{acknowledgements}
We acknowledge the referee for his/her carful reading of the paper and for suggestions that improved this work. 
We would like to deeply acknowledge R. R Chary for prompt replies and making available his IDL code which gives IR Luminosity and thereby SFR. 
We also thanks  P. Coelho for fruitful discussions, and E. S. Cypriano and C. Adami for making the previous Gemini reduced data available for this work. 
T. F. L acknowledges FAPESP for financial support through grant 2012/00578-0. L. M thanks CNPq for financial support through grant 305291/2012-2, 
and M. P. U thanks NAT/UCS for hosting while we carried out part of this research.
\end{acknowledgements}

\bibliography{refs}

\appendix
\renewcommand{\thefigure}{\thesection.\arabic{figure}}

\section{Notes on individual galaxies}
\label{notes}
Here we give more details about the sources that we used a reduced aperture to get MIPS flux density and also a merging pair of galaxy that might 
have the total flux contaminated.

\begin{itemize}

\item Source 1, 6 and 17: These source have a \texttt{I1$\_$FluxType=4}, which means that this source is real but its flux density cannot be measured accurately
and we had to use a reduced  aperture (due to nearby sources) to get the flux density and the values presented in Tab.\ref{tab_res} are lower limit values.
 Source 1 and 6: These sources are clearly about 10 sigma by eye, but there are several sources
nearby so  that we used a 2.5 pixel radius aperture. The uncertainties are based on the SHA provided values
elsewhere and that the uncertainty map is nearly constant for
the inner region. Source 17: This source is only 3.4 SNR in SHA table, thus we used reduced aperture to get flux due to
nearby source. Instead of fixed MISP aperture of 7.35 pixels, we used 4 pixels. 

 \item Source 10 (RA DEC): This is a clearly merging pair of galaxies. The total flux obtained from MIPS might not be
  due to a single galaxy.
\end{itemize}

\section{Best fits}

\begin{table*}[ht!]
\small
  \caption{We applied a robust linear regression through Bisquare weights method to obtain power-law fits, $\log(Y)=A+B\log(X)$.
  Uncertainties were determined using a Bootstrap linear fit to the data with 100 resampling.}
\begin{center}
\begin{tabular}{cccc r@{.}l  r@{.}l }
\hline
\hline
Y  & X  &  Sample & A & B & Pearson correlation coef.\\  
\hline
$\rm SFR_{\rm MIPS}$ & Proj. Distance  & Whole & 0.95 $\pm$ 0.27 & -0&59 $\pm$  0.21 & -0&34 \\
$\rm SFR_{\rm MIPS}$ & Proj. Distance &  $M_{\star} \le 1.2 \times 10^{11} M_{\odot}$ & 1.03  $\pm$ 0.17 & -0&0030 $\pm$ 0.19 & -0&0038 \\
$\rm SFR_{\rm MIPS}$ & Proj. Distance &  $M_{\star} \ge 1.2 \times 10^{11} M_{\odot}$ &  0.96 $\pm$ 0.35 & -0&97 $\pm$ 0.34  &  -0&49 \\
\hline
$\rm SFR_{\rm MAGPHYS}$ & Proj. Distance & Whole &  1.01 $\pm$ 0.17 & -0&65 $\pm$ 0.16 & -0&53 \\
$\rm SFR_{\rm MAGPHYS}$ & Proj. Distance &  $M_{\star} \le 1.2 \times 10^{11} M_{\odot}$ &  0.95 $\pm$ 0.47 & 0& 027 $\pm$ 0.35&  -0&066 \\
$\rm SFR_{\rm MAGPHYS}$ & Proj. Distance &  $M_{\star} \ge 1.2 \times 10^{11} M_{\odot}$ &   0.98 $\pm$ 0.47 & -1&29 $\pm$ 0.13  &  -0&79  \\
\hline
$\rm SFR_{\rm MIPS}$ & $M_{\star}$  & Whole & -0.95 $\pm$ 0.47 & 0&18 $\pm$ 0.13 & 0&17 \\
$\rm SFR_{\rm MIPS}$ & $M_{\star}$  & $\rm Dist \le r_{500}$ & -0.68 $\pm$ 0.21 & 0&14 $\pm$ 0.14 & 0&16 \\
$\rm SFR_{\rm MIPS}$ & $M_{\star}$ & $\rm Dist \ge r_{500}$ & -3.87 $\pm$ 0.45 & 0&45 $\pm$ 0.19 & 0&31 \\
\hline
$\rm SFR_{\rm MAGPHYS}$ & $M_{\star}$ &  Whole &-0.97 $\pm$ 0.64 & 0&19 $\pm$ 0.10  & 0&22 \\
$\rm SFR_{\rm MAGPHYS}$ & $M_{\star}$ &  $\rm Dist \le r_{500}$ &  -0.41 $\pm$ 0.37 & 0&15 $\pm$ 0.18 & 0&20 \\
$\rm SFR_{\rm MAGPHYS}$ & $M_{\star}$ &  $\rm Dist \ge r_{500}$ & -0.99 $\pm$ 0.77 & 0&18 $\pm$ 0.16 &  0&23\\
\hline
\hline
$\rm SSFR_{\rm MIPS}$ & Proj. Distance  & Whole & -9.80 $\pm$ 0.26 & -0&067 $\pm$ 0.09 & -0&03   \\
$\rm SSFR_{\rm MIPS}$ & Proj. Distance &  $M_{\star} \le 1.2 \times 10^{11} M_{\odot}$ & -9.60 $\pm$ 1.61 & -0&16 $\pm$ 0.10 & -0&10 \\
$\rm SSFR_{\rm MIPS}$ & Proj. Distance &  $M_{\star} \ge 1.2 \times 10^{11} M_{\odot}$ & -10.22 $\pm$ 0.63 & -0&33 $\pm$ 0.13 & -0&19 \\
\hline
$\rm SSFR_{\rm MAGPHYS}$ & Proj. Distance  & Whole &  -9.82 $\pm$ 1.82 & -0&060 $\pm$ 0.26 & -0&031    \\
$\rm SSFR_{\rm MAGPHYS}$ & Proj. Distance &  $M_{\star} \le 1.2 \times 10^{11} M_{\odot}$ &  -9.65 $\pm$ 2.92 & -0&25 $\pm$ 0.12& -0&10  \\
$\rm SSFR_{\rm MAGPHYS}$ & Proj. Distance &  $M_{\star} \ge 1.2 \times 10^{11} M_{\odot}$ &  -10.22 $\pm$ 0.27 & -0&32 $\pm$ 0.11& -0&33  \\
\hline
$\rm SSFR_{\rm MIPS}$ & $M_{\star}$  & Whole & -3.63 $\pm$ 0.80 & -0&57  $\pm$ 0.24 & -0&52 \\
$\rm SSFR_{\rm MIPS}$ & $M_{\star}$  & $\rm Dist \le r_{500}$ &  -0.66 $\pm$ 0.21 & -0&86 $\pm$ 0.36 &  -0&63 \\
$\rm SSFR_{\rm MIPS}$ & $M_{\star}$ & $\rm Dist \ge r_{500}$ &  -3.87 $\pm$ 1.51 & -0&54 $\pm$ 0.35 & -0&39  \\
\hline
 $\rm SSFR_{\rm MAGPHYS}$ & $M_{\star}$  & Whole &  -0.31 $\pm$ 0.16 & -0&87 $\pm$ 0.18 & -0&75 \\
$\rm SSFR_{\rm MAGPHYS}$ & $M_{\star}$  & $\rm Dist \le r_{500}$ &  -2.62 $\pm$ 1.06 & -0&62 $\pm$ 0.30 & -0&46 \\
$\rm SSFR_{\rm MAGPHYS}$ & $M_{\star}$ & $\rm Dist \ge r_{500}$ &   -0.99 $\pm$ 0.17 & -0&81 $\pm$ 0.26  & -0&74 \\
\hline
\hline
\end{tabular}
\end{center}
\label{tab_fit}
\end{table*}

\section{MAGPHYS plots}
\label{magphysfigs}

\begin{figure*}[ht!]
 \centering
\includegraphics[width=0.45\textwidth]{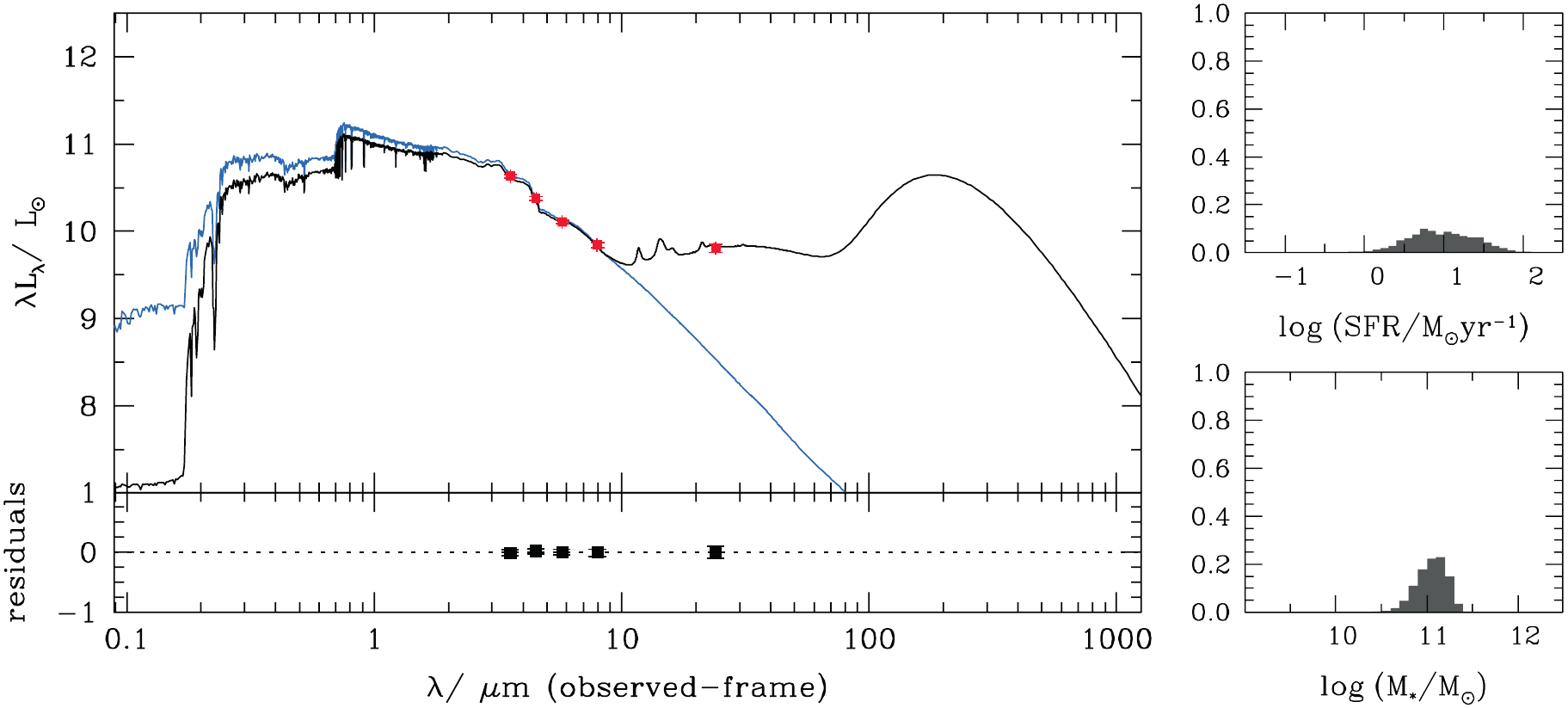} 
\includegraphics[width=0.45\textwidth]{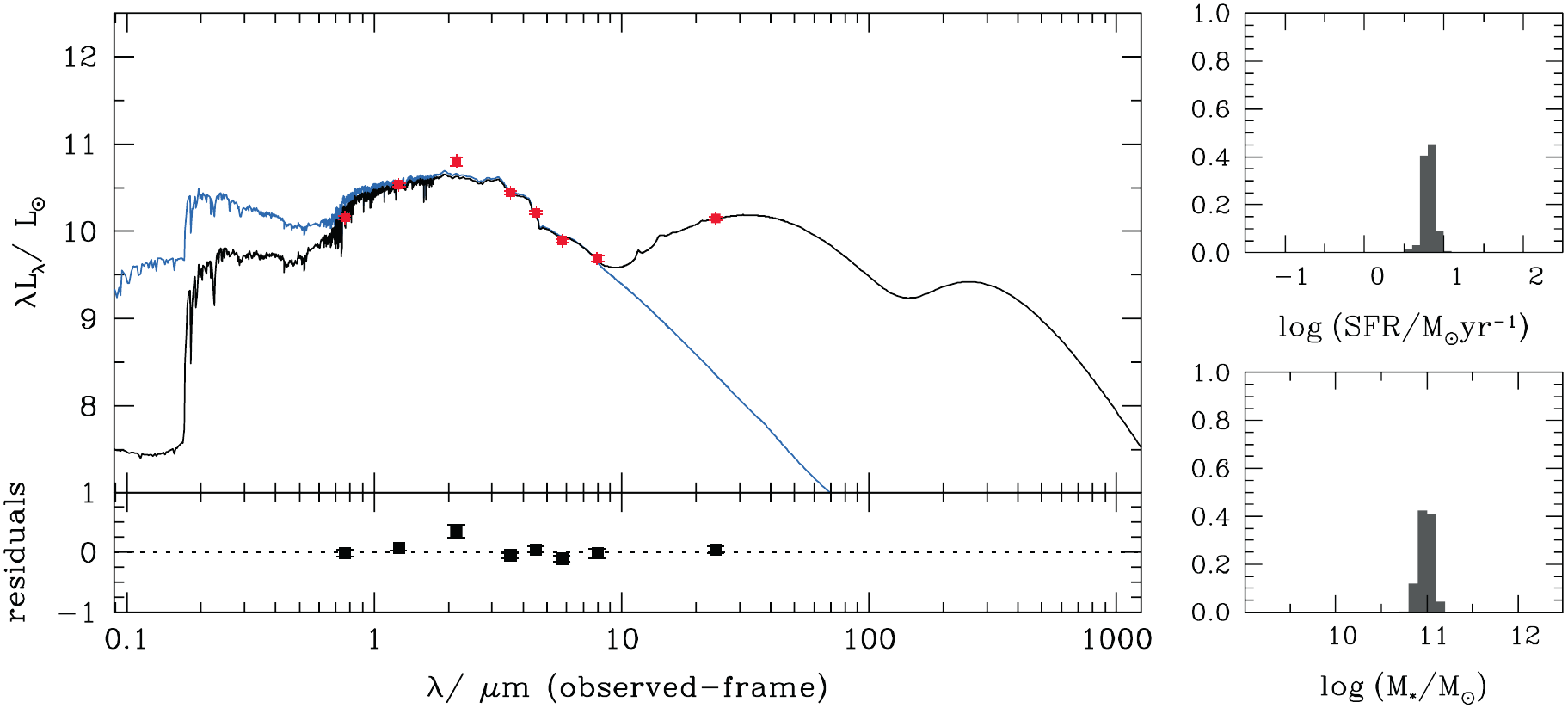} \\
\includegraphics[width=0.45\textwidth]{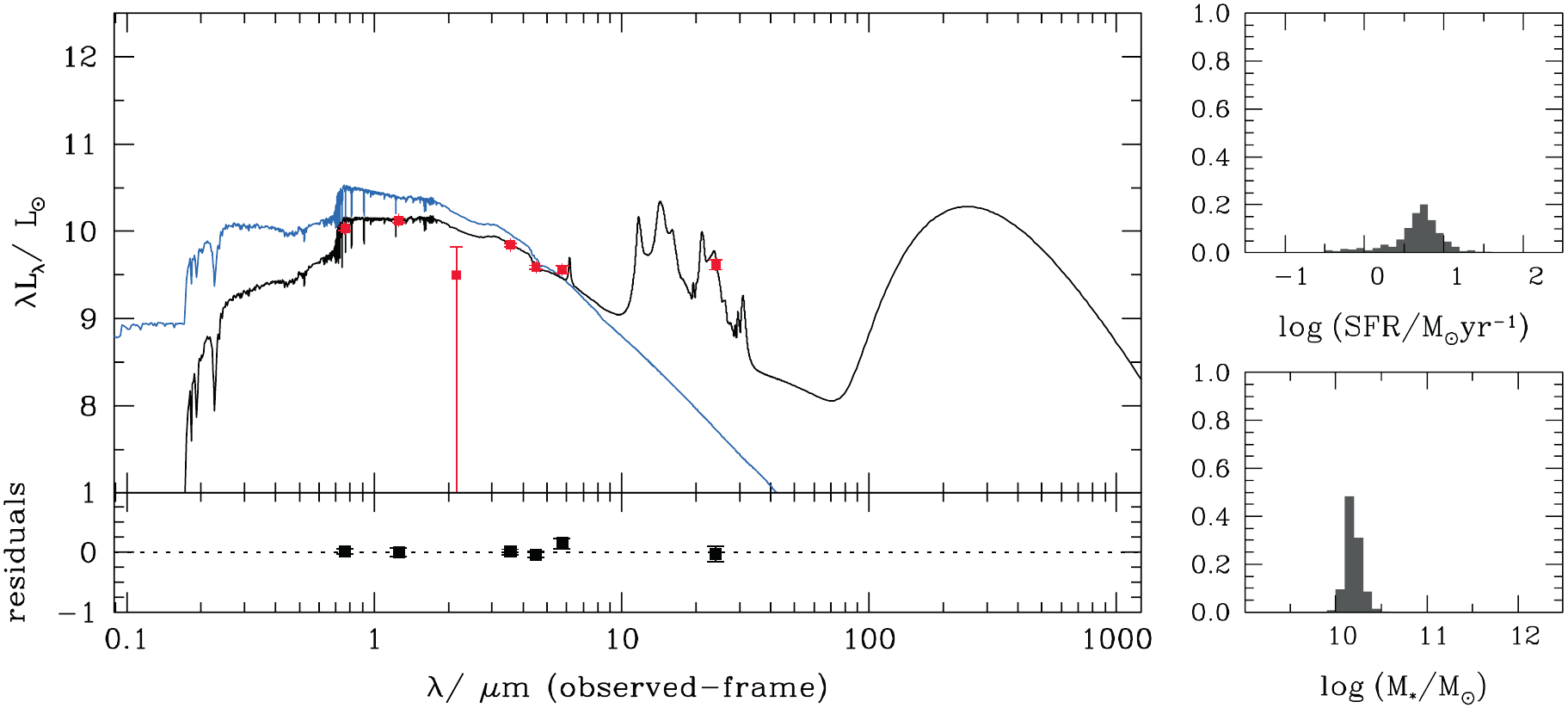}
\includegraphics[width=0.45\textwidth]{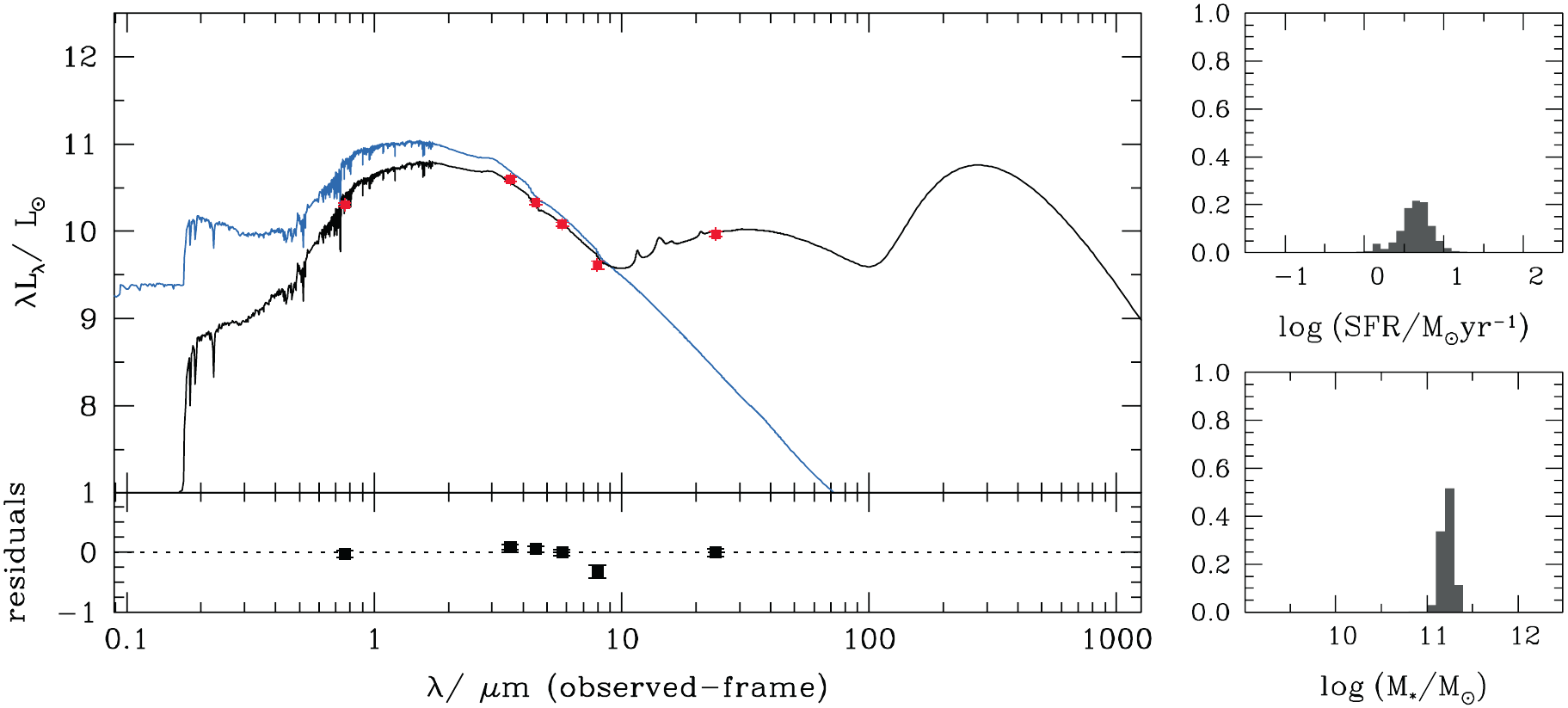} \\
\includegraphics[width=0.45\textwidth]{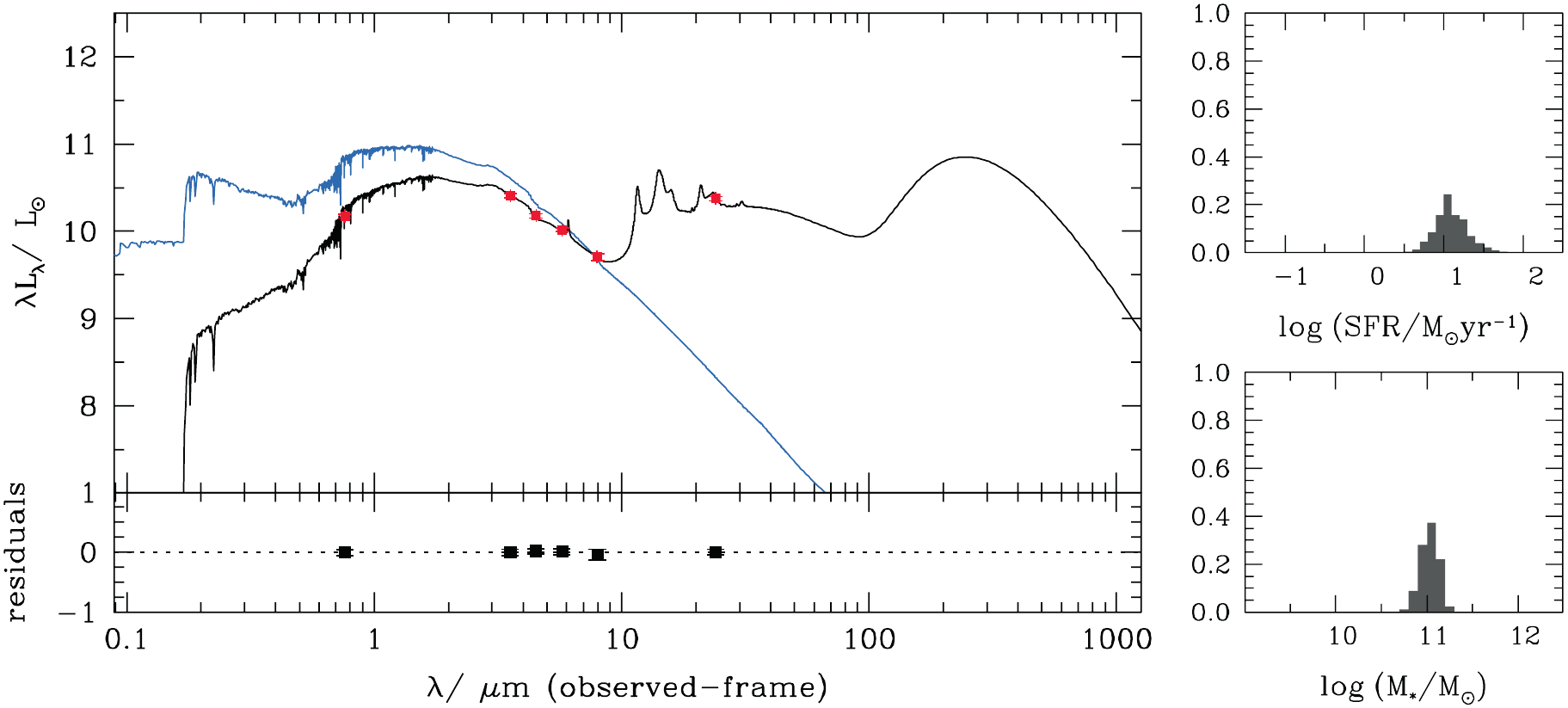} 
\includegraphics[width=0.45\textwidth]{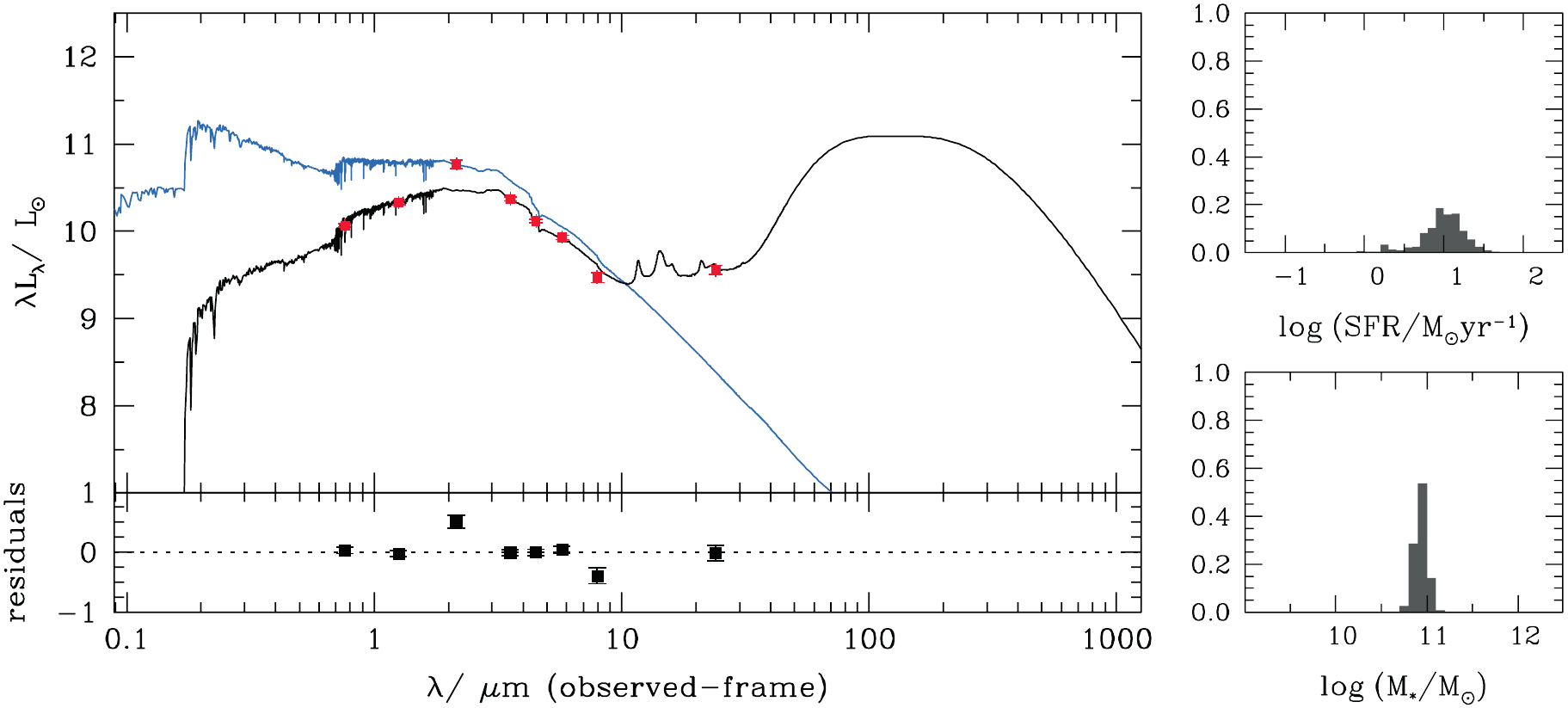}\\
\includegraphics[width=0.45\textwidth]{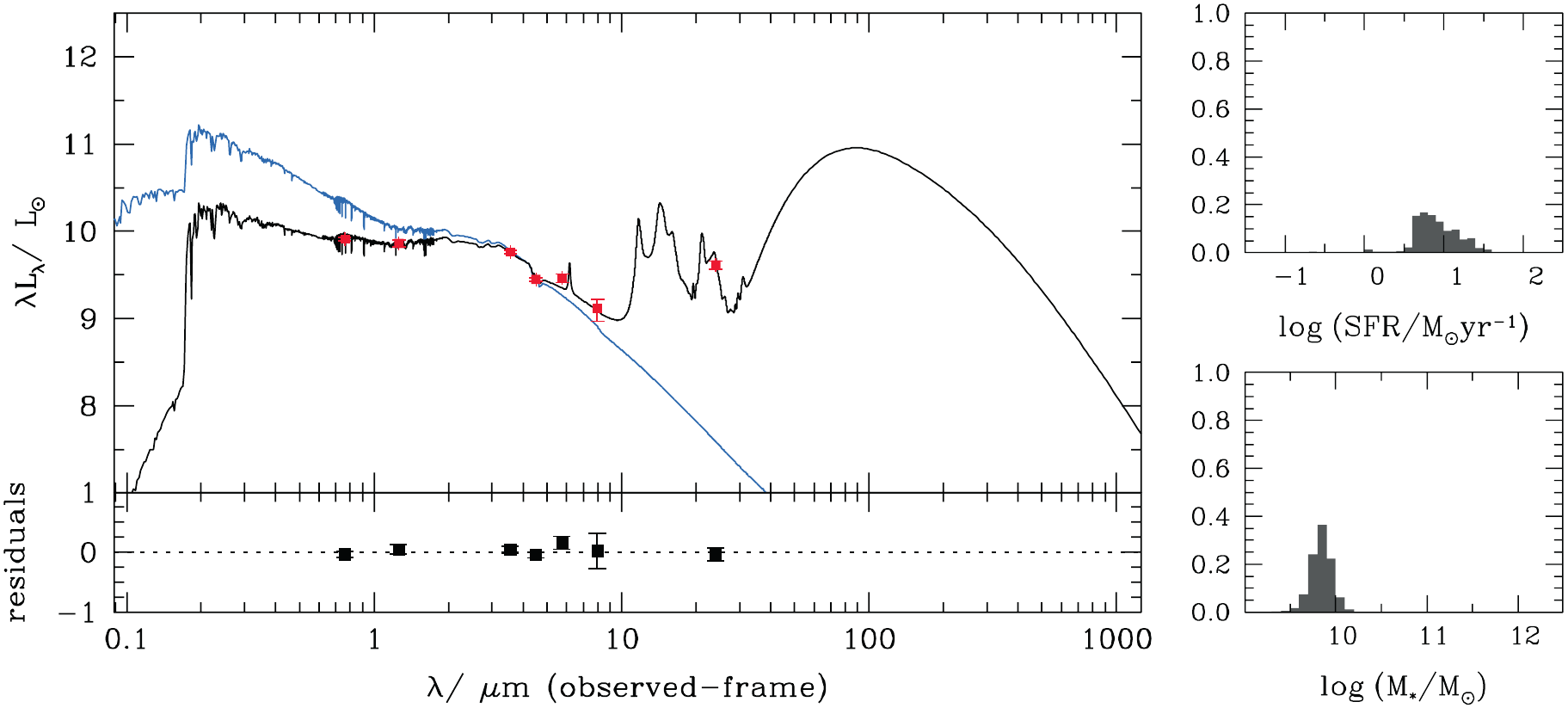} 
\includegraphics[width=0.45\textwidth]{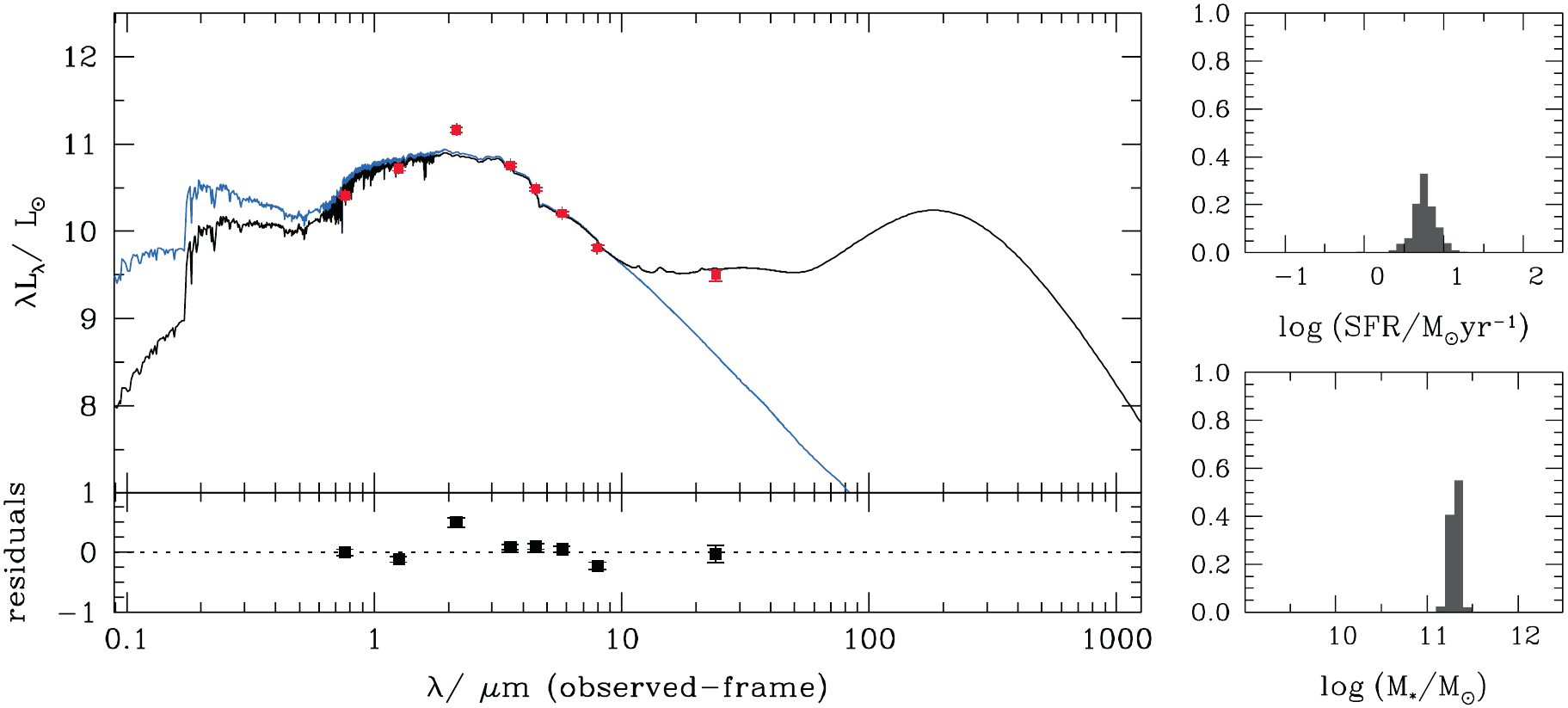} \\
\includegraphics[width=0.45\textwidth]{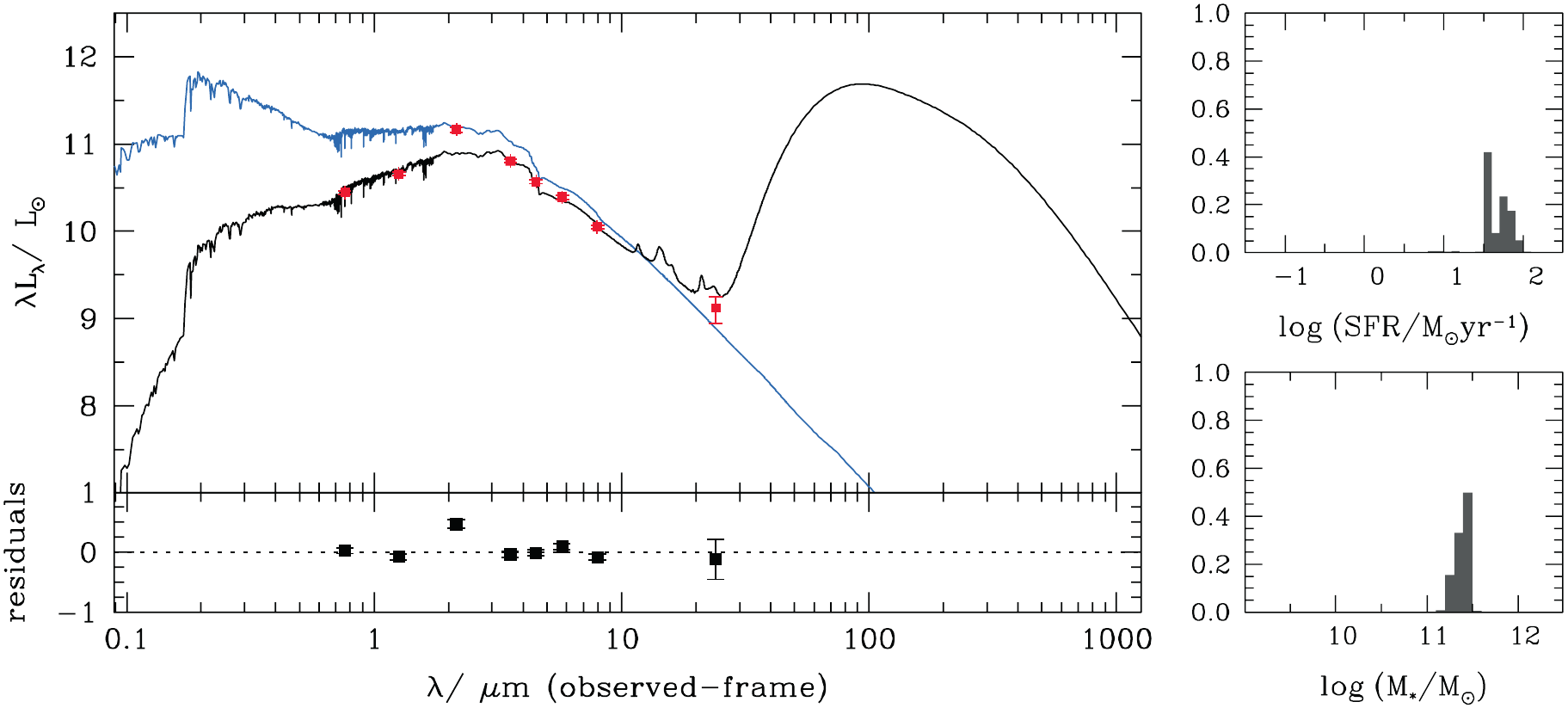} 
\includegraphics[width=0.45\textwidth]{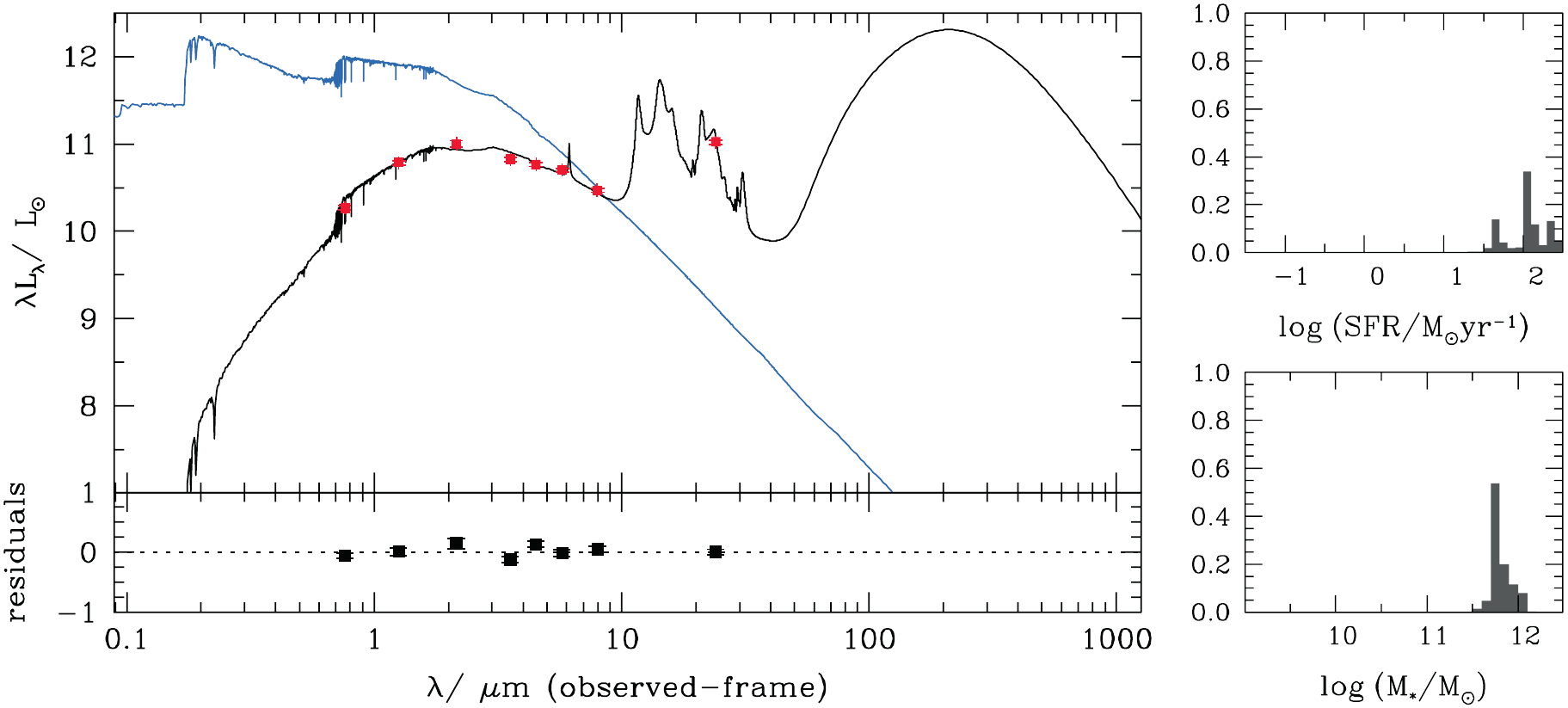} 
 \caption{Best model fits (black line) to the observed spectral energy distribution (red points) of the galaxies. In each panel, the blue line is the unattenuated stellar
 population spectrum. For each observational point, the vertical error bar indicates the measurement error. The two minor panels show the likelihood distribution of SFR (upper right panel) 
 and stellar mass (lower right panel) derived from fits to the observed spectral energy distribution. 
 From top to bottom and left to right, galaxies are in RA order, as presented in Tab.\ref{tab_res}, 
 with source codes: 1, 2, 3, 4, 5, 6. 7, 8, 9 and 10. }
  \label{fig_magphys}
\end{figure*}

\begin{figure*}[ht!]
 \centering
\includegraphics[width=0.45\textwidth]{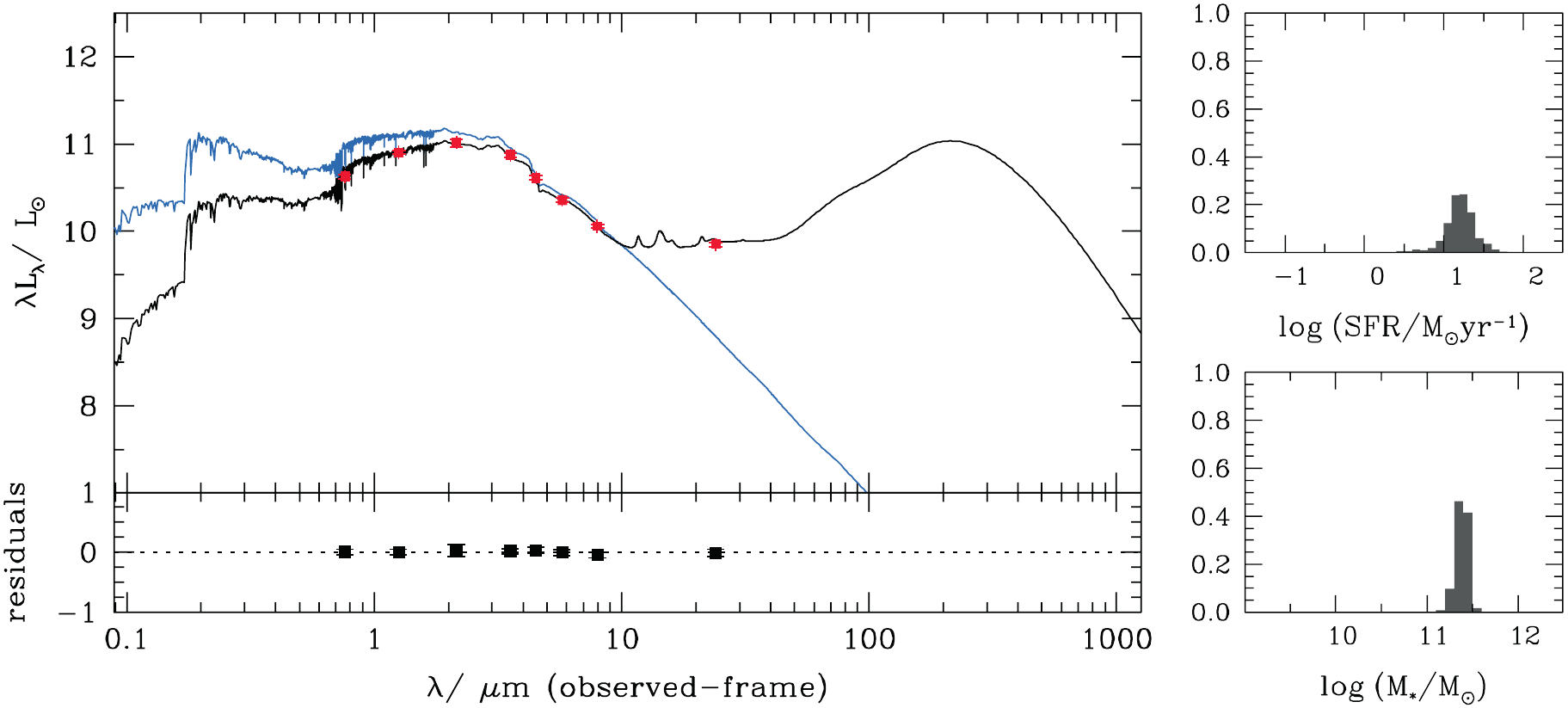} 
\includegraphics[width=0.45\textwidth]{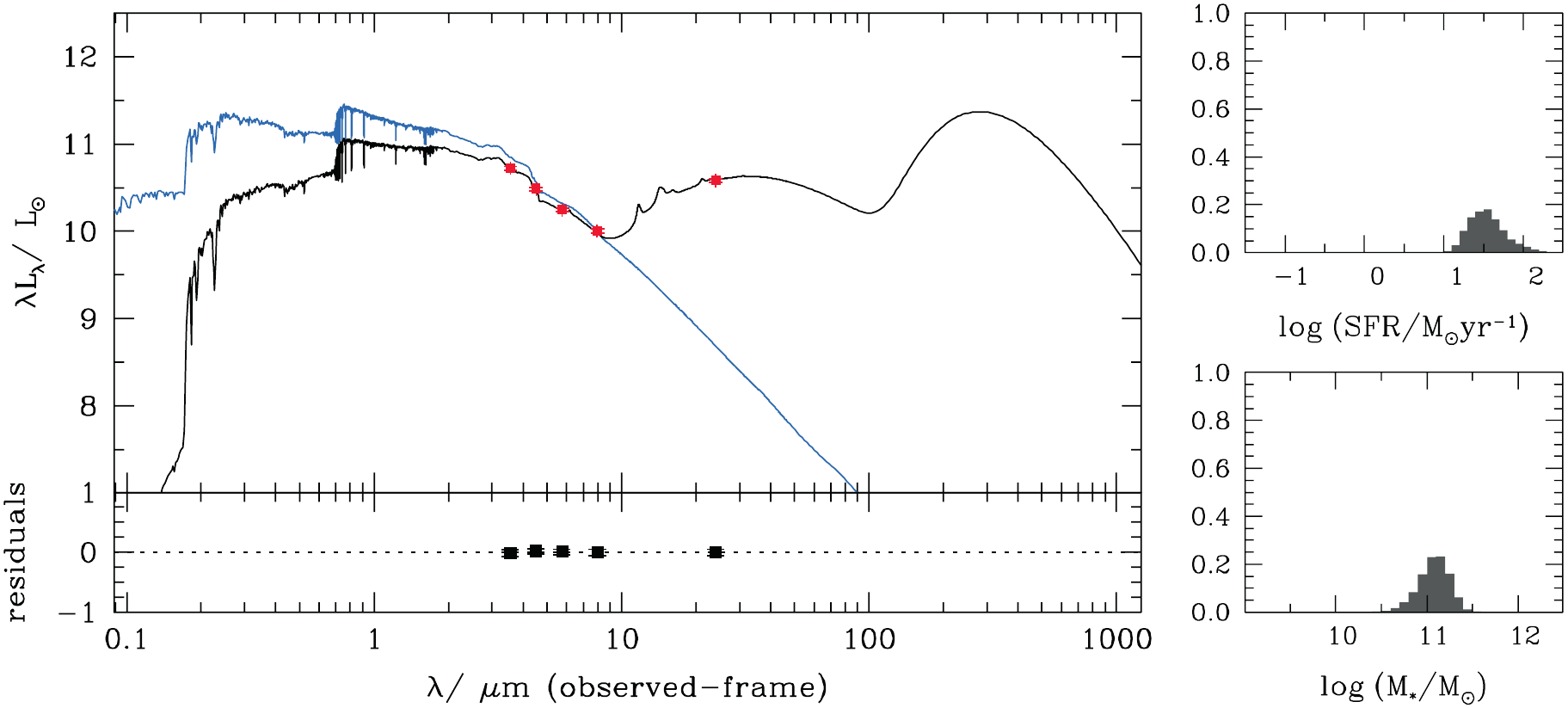}\\ 
\includegraphics[width=0.45\textwidth]{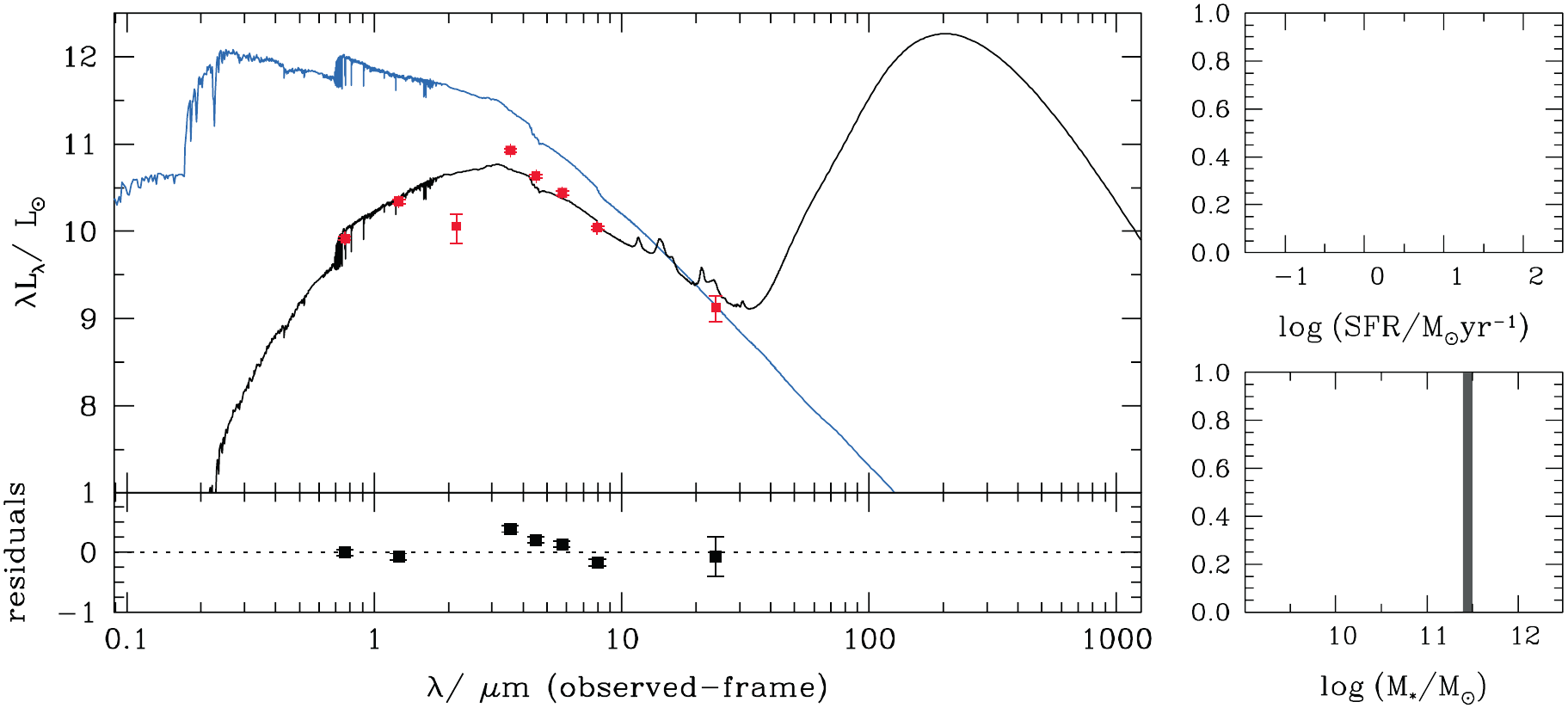} 
\includegraphics[width=0.45\textwidth]{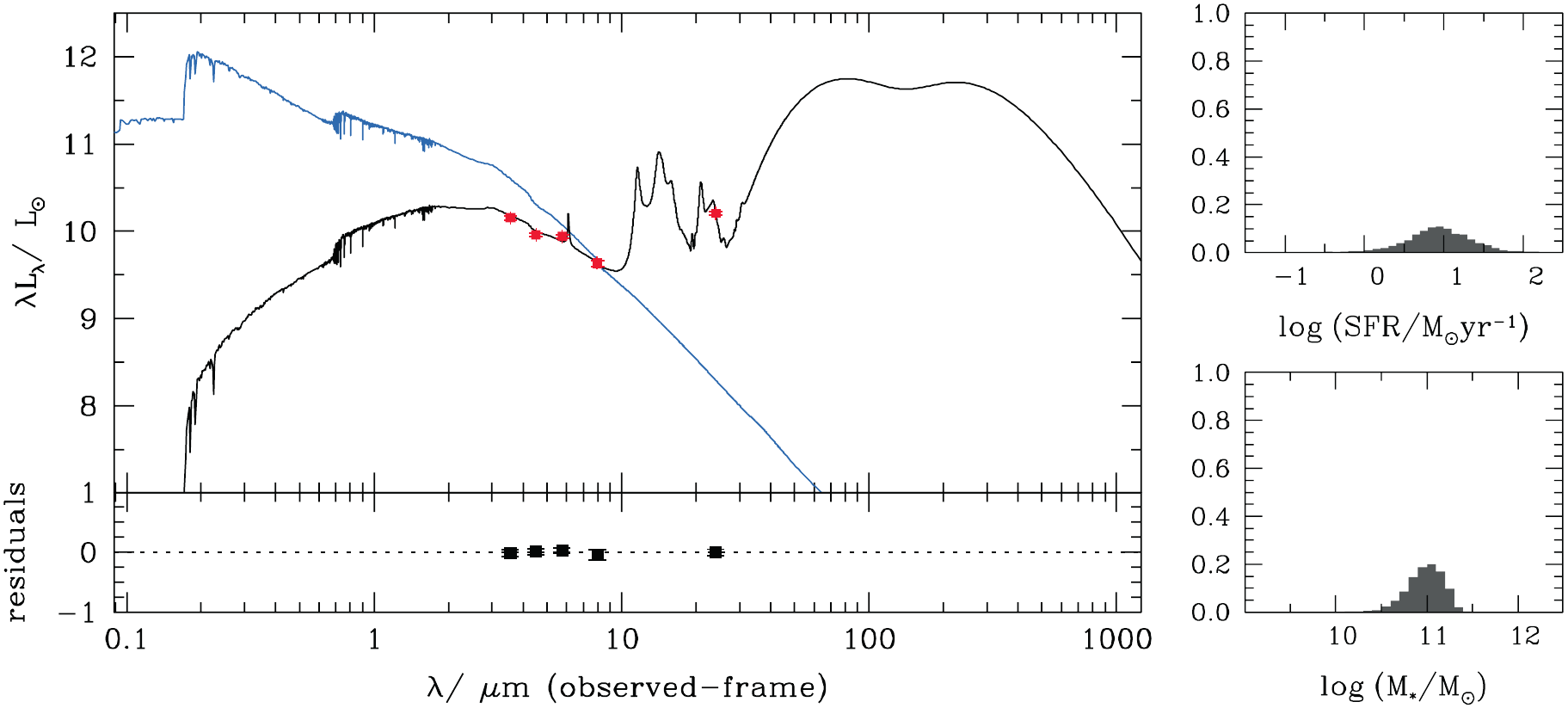} \\
\includegraphics[width=0.45\textwidth]{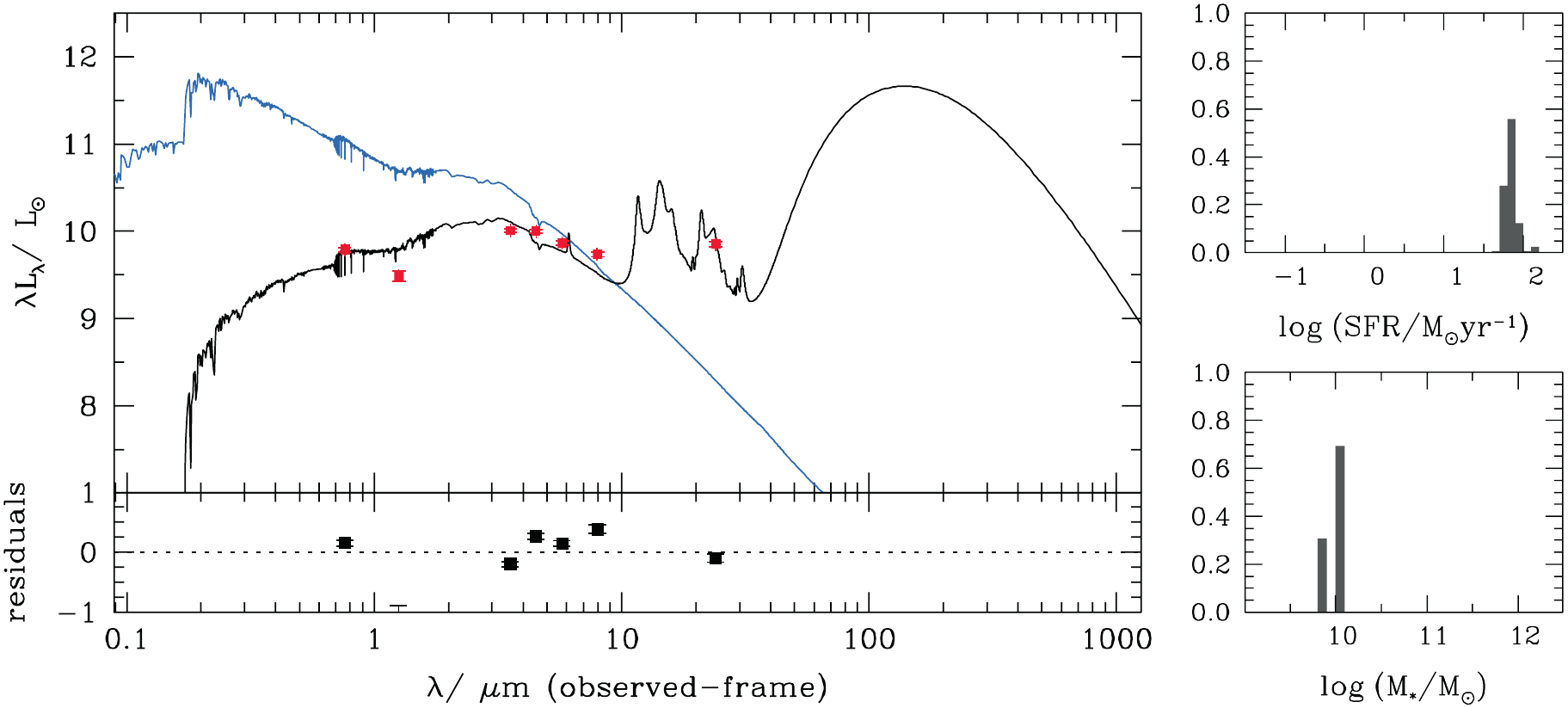} 
\includegraphics[width=0.45\textwidth]{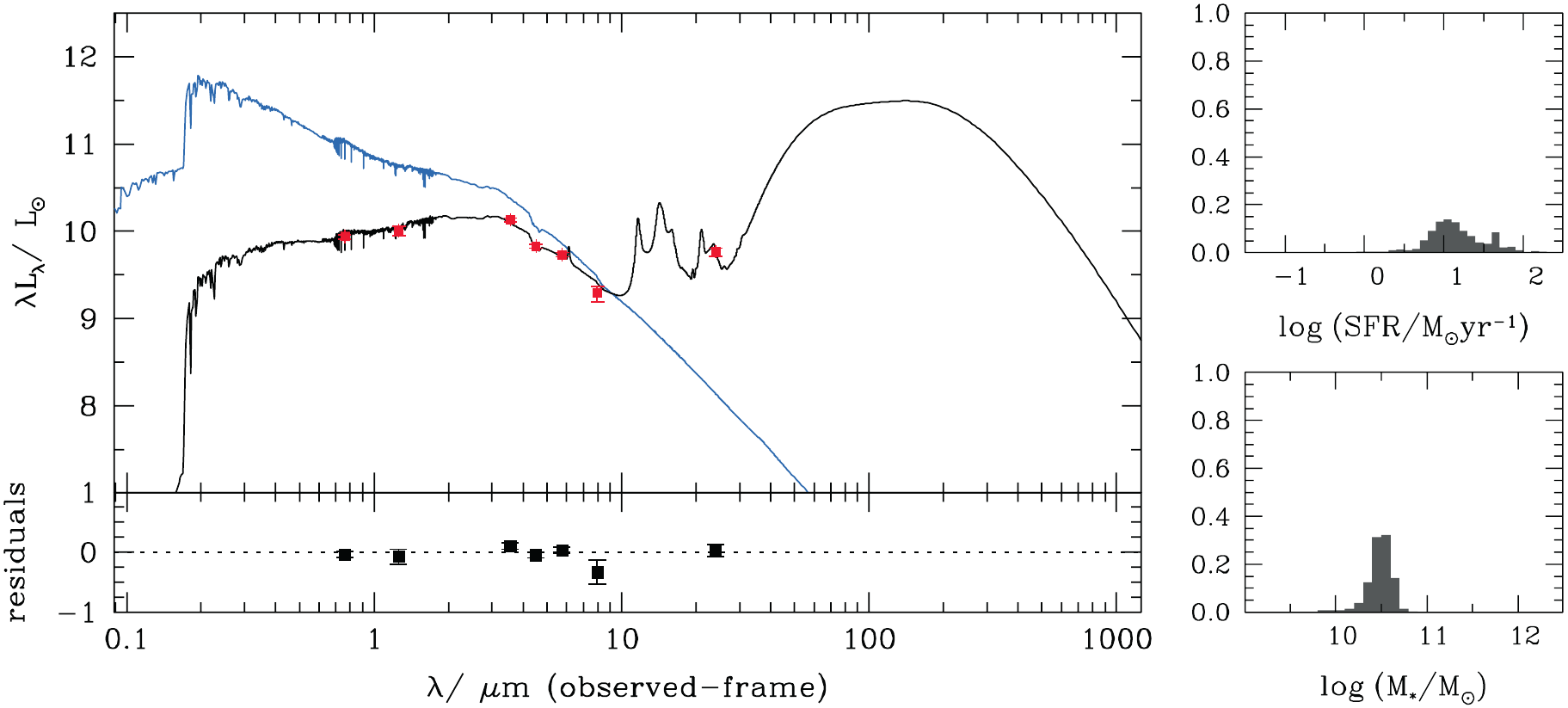} \\
\includegraphics[width=0.45\textwidth]{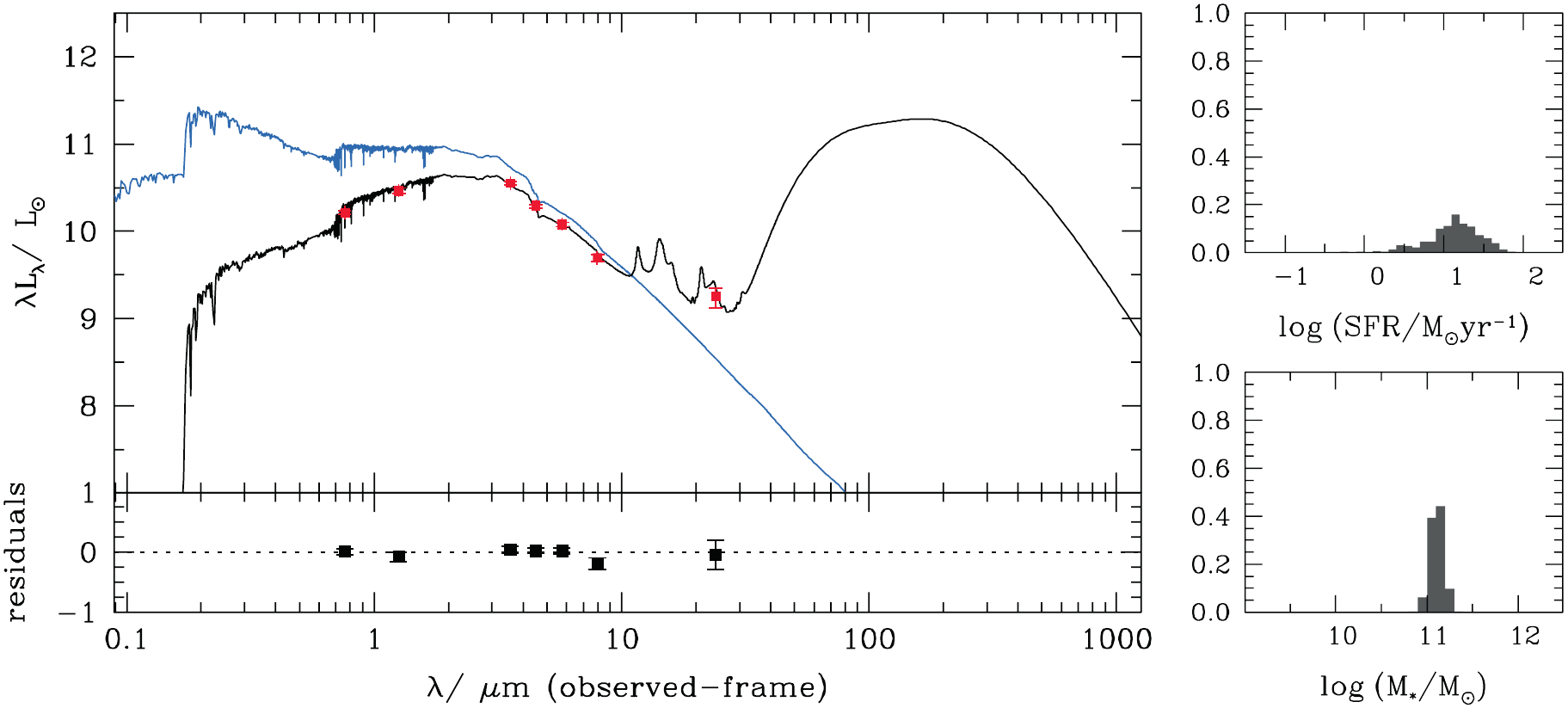}  
\includegraphics[width=0.45\textwidth]{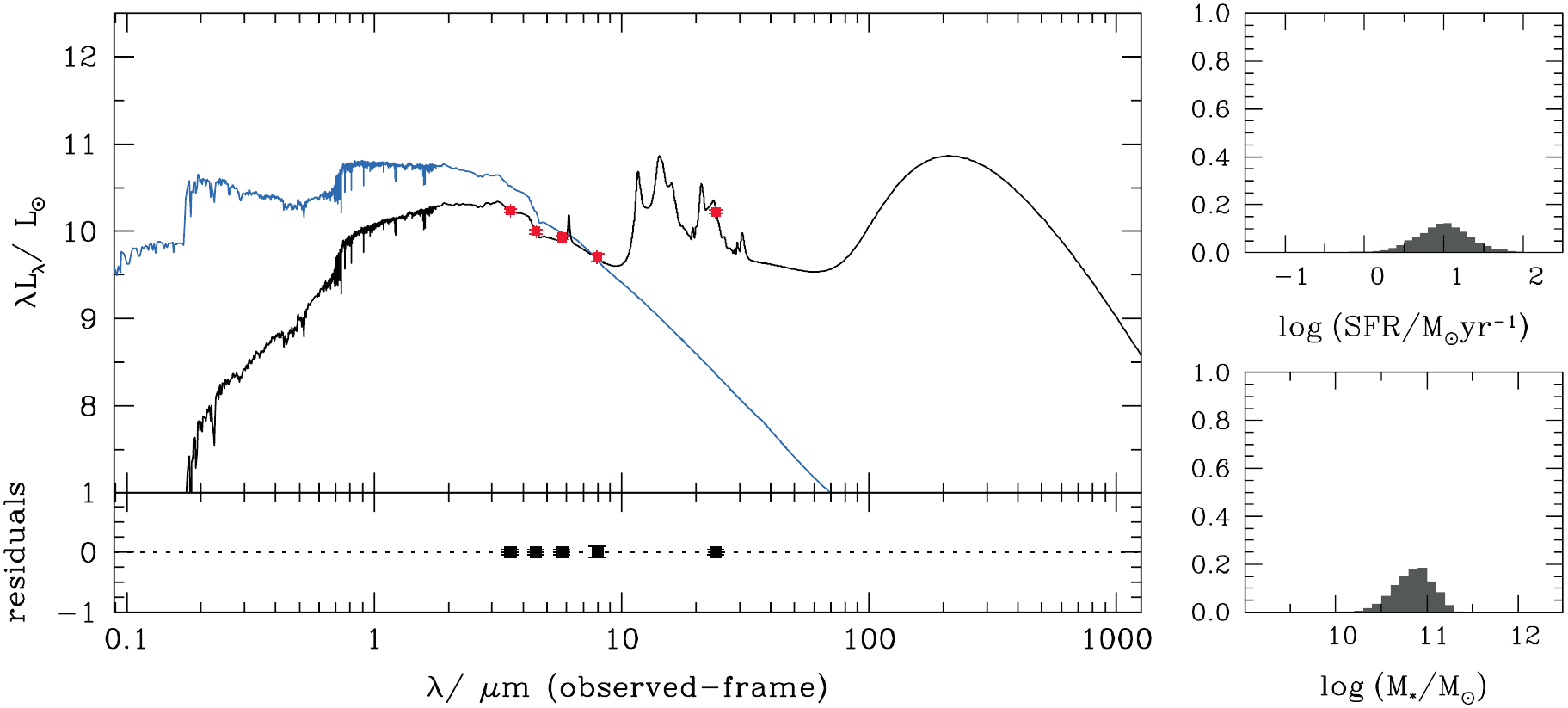}\\
 \caption{Best model fits (black line) to the observed spectral energy distribution (red points) of the galaxies. In each panel, the blue line shows the unattenuated stellar
 population spectrum. For each observational point, the vertical error bar indicates the measurement error. The two minor panels show the likelihood distribution of SFR (upper right panel) 
 and stellar mass (lower right panel) derived from fits to the observed spectral energy distribution.
 From top to bottom and left to right, galaxies are in RA order, as presented in Tab.~\ref{tab_res},
with source codes: 11, 12, 13, 14, 15, 16, 17 and 18.}
  \label{fig_magphys2}
\end{figure*}

\end{document}